\documentclass[]{aa} 
\usepackage{txfonts}
\usepackage{natbib}
\usepackage{graphicx}
\usepackage{amsmath}
\usepackage{lscape}
\usepackage{longtable}
\usepackage{xspace}
\bibpunct{(}{)}{;}{a}{}{,} 

\begin{document}

\title{Reanalyzing the visible colors of Centaurs and KBOs:
}
\subtitle{what is there and what we might be missing.}
\author{Nuno Peixinho\inst{1}
\and Audrey Delsanti\inst{2}
\and Alain Doressoundiram\inst{3}
}
\institute{Unidad de Astronom\'{\i}a, Fac. de Ciencias B\'asicas, Universidad de
Antofagasta, Avda. U. de Antofagasta 02800, Antofagasta, Chile\\ 
\email{nuno.peixinho@uantof.cl}
\and Aix Marseille Universit\'e, CNRS, LAM (Laboratoire d'Astrophysique de Marseille) UMR 7326, 13388 Marseille, France\\ 
\email{Audrey.Delsanti@lam.fr}
\and Observatoire de Paris, Site de Meudon, 5 place Jules Janssen, 92190 Meudon, France\\ 
\email{Alain.Doressoundiram@obspm.fr}
}
\date{Received 30 November 2014 / Accepted 6 February 2015}


\abstract {
Since the discovery of the Kuiper Belt, broadband surface colors were thoroughly studied as a first approximation to the object reflectivity spectra. Visible colors (BVRI) have proven to be a reasonable proxy for real spectra, which are rather linear in this range. 
In contrast, near-IR colors (JHK bands) could be misleading when absorption features of ices are present in the spectra.
Although the physical and chemical information provided by colors are rather limited, broadband photometry remains the best tool for establishing the bulk surface properties of KBOs and Centaurs.
In this work, we explore for the first time general, recurrent effects in the study of visible colors that could affect the interpretation of the scientific results: i) how a correlation could be missed or weakened as a result of the data error bars, ii) the ``risk'' of missing an existing trend because of low sampling, and the possibility of making quantified predictions on the sample size needed to detect a trend at a given significance level --- assuming the sample is unbiased, iii) the use of partial correlations to distinguish the mutual effect of two or more (physical) parameters, and iv) the sensitivity of the ``reddening line'' tool to the central wavelength of the filters used. 
To illustrate and apply these new tools, we have compiled the visible colors and orbital parameters of about 370 objects available in the literature 
--- assumed, by default, as unbiased samples --- and carried
out a traditional analysis per dynamical family.
Our results show in particular how a) data error bars impose a limit on the detectable correlations regardless of sample size and that therefore, once that limit is achieved, it is important to diminish the error bars, but it is pointless to enlarge the sampling with the same or larger errors; b) almost all dynamical families still require larger samplings to ensure the detection of correlations stronger than $\pm$0.5, that is, correlations that may explain $\sim$25$\%$ or more of the color variability; 
c) the correlation strength between (V-R) vs. (R-I) is systematically lower than the one between (B-V) vs. (V-R) and is not related with error-bar differences between these colors;  
d) it is statistically equivalent to use any of the different flavors of orbital excitation or collisional velocity parameters regarding the famous color-inclination correlation among classical KBOs --- which no longer appears to be a strong correlation --- whereas the inclination and Tisserand parameter relative to Neptune cannot be separated from one another; and 
e) classical KBOs are the only dynamical family that shows neither (B-V) vs. (V-R) nor (V-R) vs. (R-I) correlations. It therefore
is the family with the most unpredictable visible surface reflectivities.}

\maketitle


\section{Introduction}

Kuiper Belt objects (KBOs), also known as Trans-Neptunian objects (TNOs), are a large population of 
icy bodies orbiting the Sun with semi-major axes greater than
that of Neptune. 
They are considered as the least altered objects in the solar system. 
Their existence had been hypothesized by \cite{Leo30}, \cite{Edg43, Edg49}, and \cite{Kui51} --- 
and the expression Edgeworth-Kuiper Belt is also used in some works --- but they were only observed in 1992 by \cite{JewLuu93}. 
Currently, there are more than 1700 objects identified, $\sim 4.5\%$ are binary or multiple systems, 
and there are probably $\sim 200\,000$ objects larger then $100\,km$ in diameter 
\citep{Petit+2011}. 

KBOs subdivide into several dynamical families, although there is no strict definition regarding the boundaries between 
most of them \cite[see][for a review]{Gladman+2008}.
Centaurs are, presumably, former KBOs that evolved into inner orbits, that is, both their semi-major 
axes and perihelia lie between the orbits of Jupiter and Neptune \citep{Fernandez1980,LevDun97} ---  
although when using a less strict dynamical definition of Centaurs, some may have an Oort Cloud origin \citep{Emelyanenko+2013, Fouchard+2014b}.
Every $\sim 125\, yr$ a new Centaur escapes from the Kuiper Belt into a 
short-lived chaotic orbit, with an average half-live of $\sim 2.75\,Myr$ --- although some may last for millions of years --- before becoming a 
Jupiter-family comet (JFC) or being 
re-injected into the outer solar system \citep[e.g.,][]{Horner+04b}. Given this close link, Centaurs have usually been studied as one more family of KBOs. 

Since the first analysis by \cite{1996AJ....112.2310L} of a sample of three Centaurs and nine KBOs, any relation between the surface properties observed at different wavelengths has been searched
for, such as different colors. These relations might be 
among the surface properties themselves or between the surface properties and the orbital parameters. 
Although correlation analysis does not allow us to predict specific values of unmeasured properties by knowing others, 
nor to draw conclusions about cause and effect relationships, it is the most elementary tool for detecting and quantifying which variables or properties covary with each other, which provides a first insight into the phenomena we are observing. 

Many works reported trends, patterns, and correlations between different surface colors 
and between those and the orbital properties of these objects \citep[see review][]{2008ssbn.book...91D}. 
While several teams had their own surveys --- for example, the 
Tegler, Romanishin, and Consolmagno survey \citep[e.g.,][]{1998Natur.392...49T,2000Natur.407..979T,2003Icar..161..181T,2003ApJ...599L..49T},
the ESO Large Program on Centaurs and TNOs \citep[e.g.,][]{2001A&A...380..347D,2002A&A...395..297B,2004A&A...417.1145D,2004Icar..170..153P,2006AJ....131.1851D},
the Meudon Multicolor Survey (2MS), \citep[e.g.,][]{1999Icar..142..476B,2000AJ....120..496B,2001Icar..154..277D,2002AJ....124.2279D,2005Icar..174...90D,2007AJ....134.2186D},
and the Second ESO Large Program \citep[e.g.,][and references therein]{2009A&A...493..283D,2010A&A...510A..53P} ---
focusing on the analysis of their own datasets for methodological reasons, 
\cite{HaiDel02} made the first thorough analysis of a large compilation of all published data, 
known as the MBOSS database, which has recently been updated \citep{Hainaut+2012}. 

Although the physical and chemical information provided by the visible colors of these objects is rather limited, the analysis of visible colors alone 
has lead to very rich debates on the Kuiper Belt science.

Given the narrow temperature differences in the Kuiper Belt region, KBOs were initially presumed to have a primordial homogeneous composition, 
and the first modeling works attempted to explain the visible surface color diversity as a result of a competition 
between a reddening space-weather effect of irradiation on an ice-rich surface layer and a bluishing collisional resurfacing by buried 
non-irradiated layers \citep{1996AJ....112.2310L}. 
A strong correlation of color vs. orbital inclination detected among the dynamic family of classical KBOs \citep{2002ApJ...566L.125T}
led to a further development of collisional resurfacing models, accounting for differently irradiated 
subsurface layers \citep{GilH02}, or for collision-triggered cometary activity 
and size-dependent resurfacing \citep{2004A&A...417.1145D}. While these models had difficulties in explaining the detection, or non-detection, of 
several surface properties and correlations between these and the orbital properties \citep[e.g.,][]{2001AJ....122.2099J, TheDor03, 2004A&A...417.1145D}, the dynamical model of 
\cite{Gomes03} suggested that high-inclination KBOs formed closer to the Sun migrate afterward into their current orbits. 
This opened a new window for the possibility of original compositional differences among KBOs \citep[e.g.,][]{2003ApJ...599L..49T}.

Simultaneously, a debate on the global existence, or absence, of only two distinct surface color groups of Centaurs and KBOs \citep{1998Natur.392...49T,2003Icar..161..181T}
led to the conclusion that only Centaurs seemed to be separated into two distinct groups, whereas KBOs presented a somewhat 
continuous distribution of colors \citep{Peix+03L}, and current data suggest that the existence of two distinct groups of objects 
is related to their sizes and not to their dynamical families \citep{2012A&A...546A..86P, FraserBrown2012}.
 
Dynamical models indicate that the giant planets have migrated throughout the history of the solar system, 
sculpting the structure of the current Kuiper Belt \cite[e.g.,][and references therein]{Levison+2011, Batygin+2011}. 
Although there is still a debate on the details of this process, it became more plausible that 
the Kuiper Belt is composed by a mixture of bodies formed at distinct heliocentric distances and, therefore, 
their distinct surface properties would be primarily caused by primordial compositional differences \citep[e.g.,][]{Brown+2011}.
The most recent analysis of albedos vs. visible colors of Centaurs and KBOs identifies 
a surface color and albedo separation that is also evidence for such primordial compositional differences \citep{Lacerda+2014}.
Furthermore, latest laboratory works show that it is possible to reproduce a wide range of surface colors of 
Centaurs and KBOs with an appropriate combination of initial albedos, collisional evolution, and space weathering, 
implying that it is unlikely that a straightforward dependance between the dynamical properties and the 
surface properties exists \citep{Kanuchova+2012}. Thus, it comes as no surprise 
that the two taxonomy schemes  proposed, so far, identify several groups with distinct surface properties \citep{Barucci+2005,2010A&A...510A..53P,FraserBrown2012,DalleOre+2013}, failing to detect any clear dependance between the dynamical properties and the 
surface properties, although the proportions of taxonomic families vary among the dynamical families. 

With the growing interest on detailed spectroscopy studies in the near-IR, where several ice absorption bands can be detected ,
as well as with the existence of an apparently large database of visible colors of our objects, it might be perceived as 
irrelevant to continue photometric studies of KBOs and their progeny. However, spectroscopic studies are 
still limited to fewer than 100 objects (i.e., $\sim 5\%$ of known objects) and are extremely biased toward only the brightest objects, 
whereas visually all detected objects might be studied photometrically, and, as aforementioned, it is currently known that 
smaller (fainter) objects do possess a peculiar bimodality of surfaces that is not seen among the larger (brighter) objects. 
The question is, nonetheless, pertinent, as in theory one does not need to study all the population but only a representative sample of it. 
The problem is that the more subtle the effect one seeks, the larger the sample required. 

In this work we exclusively focus on studying the correlations between the visible surface colors and the orbital parameters 
for each traditional dynamical family of KBOs and Centaurs of our own compilation of visible colors available in the literature, 
assumed to be unbiased samples of each family, by default.
We analyze and discuss for the first time,what we might be missing as a function of the available sample 
sizes for each dynamical family, and also as a function of the observational error-bars. 
For an analysis of the (statistical) distributions of the surface properties of each dynamical family see \cite{Hainaut+2012}. 


\section{Data sample}
\label{sec:data_sample}

Orbital elements were gathered from the Asteroid Orbital Elements Database, 
\verb+astorb.dat+ \footnote{ftp://ftp.lowell.edu/pub/elgb/astorb.dat.gz, with epoch 20120420}, maintained by the Lowell Observatory  
based on astrometric observations by the Minor Planet Center. Orbital inclinations $i$ are relative to the ecliptic 
and not to the so-called Kuiper Belt plane. 

We used the classification scheme suggested by \cite{LykMuk07}, including their analysis of objects located in 
the mean motion resonances (MMR) with Neptune, following a ten-step algorithm:

\begin{enumerate}

\item in $3:2$ MMR with Neptune $\Rightarrow$ Plutino
\item in other MMR with Neptune $\Rightarrow$ other resonant
\item $q<a_J \Rightarrow$ not analysed
\item $q>a_J \wedge a<a_N \Rightarrow$ Centaur
\item $a_J < q < a_N \wedge a\geqslant a_N \Rightarrow$ scattered disk object (SDO)
\item $a_N < q \leqslant 37 \,AU  \Rightarrow$ scattered disk object (SDO)
\item $q\geqslant 40 \,AU \wedge a\geqslant 48 \,AU  \Rightarrow$ detached KBO (DKBO)
\item $37 \,AU \leqslant q \leqslant 40 \,AU  \Rightarrow$ scattered or detached KBO (SDKBO)
\item $i<5^{\circ} \wedge \{ \,[ q\geqslant 37 \,AU \wedge ( 37 \,AU \leqslant a \leqslant 40 \,AU) ] \vee [q\geqslant 38 \,AU \wedge (42 \,AU \leqslant a \leqslant 48 \,AU)]\,\} \Rightarrow$ cold classical KBO (cCKBO)
\item $i\geqslant5^{\circ} \wedge q\geqslant 37 \,AU \wedge  ( 37 \,AU \leqslant a \leqslant 48 \,AU ) \Rightarrow$ hot classical KBO (hCKBO).

\end{enumerate}

We are aware that there are more complex classification schemes
that may be more rigorous, but the boundaries between families do 
not change significantly.  We chose this one for its computational simplicity, which provides a dynamical classification for all objects in our sample, whereas dynamical simulation-based classifications are only available for a limited number of objects \citep[e.g.,][]{Gladman+2008}.

The so-called Haumea family \citep{Brown+2007, 2010A&A...511A..72S} --- $(136108)$ Haumea, $(24835)$ 1995SM$_{55}$, 
$(55636)$ 2002TX$_{300}$, $(86047)$ 1999OY$_{3}$, $(19308)$ 1996TO$_{66}$, $(120178)$ 2003OP$_{32}$, $(145453)$ 2005RR$_{43}$, 
2003SQ$_{317}$, 
2003UZ$_{117}$, and 2005CB$_{79}$ --- are extremely peculiar objects, therefore we removed them from the statistical analysis. We also removed the two objects in retrograde orbits (2008KV$_{42}$, 2008YB$_{3}$).
For a total of 366 objects in our database, we carried out a global analysis, that is, all objects except Haumea family and retrograde orbits, of $n=354$ objects, 
calling it All$^{*}$; another analysis also removed Centaurs, which we call KBOs$^{*}$; a third analysis according to dynamical family; and finally, an analysis for which we combined some families. 

We compiled the visible colors ($BVRI$ broadband filters or equivalent filters transformed to that system) available for 366 objects 
(see Table \ref{tab:colors}) and fully trusted their published error-bars. For each input paper and each object or observation, we computed the reflectance spectrum using 
Eq. (3) from \cite{2001A&A...380..347D}, when two or more filters were available. The resulting spectra were manually checked and obviously 
deviating data from a given filter were removed from the dataset. Color indexes were computed within one given epoch, which led to colors 
obtained from {\it \textup{simultaneous}} photometry (the different bands were observed over a maximum time-span of two hours). Then the average 
colors indexes and their one-sigma errors from different papers and epochs were computed for each object using 
Eqs. (1) and (2) from \cite{HaiDel02}, which provided more accurate estimates when multiple measurements are available. 
The absolute magnitude in $R$ band ($H_R$) was computed for each object and epoch whenever an $R$-band magnitude was available, using 
$H_R = R - 5 \log (r \Delta)$, 
where $R$ is the $R$-band magnitude and $r$ and $\Delta$ the helio- and geocentric distances at the time of observations, respectively.
Different values for a given object were also averaged using Eqs. (1) and (2) from \cite{HaiDel02}. 
We did not correct for any phase effect, as already discussed in \cite{2012A&A...546A..86P}; see Table \ref{tab:colors}.


\section{Statistical methods}
\label{sec:stat_methods}


\subsection{Statistical tests}
\label{sec:stat_tests}

\begin{itemize}


\item {\bf Spearman-rank correlation accounting for data error-bars:}
We analyzed the possible existence of correlations between surface colors of 
Centaurs and KBOs and orbital parameters using the Spearman-rank correlation coefficient 
$\rho$ \citep{Spe04}. The significance of $\rho$ was estimated using $t=\rho\,[(n-2)/(1-\rho^2)]^{1/2}$ , which 
obeys approximately a Student t-distribution 
with $n-2$ degrees of freedom, where $n$ is the number of data-points \citep{kendall_smith1939}.

To take into account the effect of observational errors, we considered each data-point not 
as fixed, that is, exact, but as a Gaussian distribution centered on the measured value  with  a
standard deviation equal to the corresponding error-bar. Then, we extracted 1000 new 
random data-sets in which each data-point followed its own Gaussian probability distribution. 
Since the Spearman $\rho$ -distribution is not Gaussian, we 
used the Fisher transformation $\zeta({\rho})= \frac{1}{2} \ln \big(\frac{1+\rho}{1-\rho}\big) = \arg \tanh(\rho)$ 
--- \cite{Fisher1915} ---, 
determined the mean $\langle \zeta_{\rho} \rangle$, that is, the most
probable $\zeta_{\rho}$ value, the ${+\sigma_{\zeta}}/{-\sigma_{\zeta}}$, and transformed it back to $\rho$-values 
using $\rho_\zeta=\tanh(\zeta_{\rho})$, obtaining the most probable correlation value $\langle \rho_\zeta \rangle$, 
and the shortest interval containing $68.2\%$ of the distribution, which is the equivalent of a Gaussian 
$\pm1\sigma$ limits, which in this case are not necessarily symmetrical. We denote this as 
${+\sigma_{e}}/{-\sigma_{e}}$ \citep[see][for more details]{2004Icar..170..153P}. 

This method yields the confidence interval of the most probable $\rho$ of our sample, 
but does not yield the confidence interval for the correlation value of the parent population. 
For that we made 1000 bootstrap extractions from our data-points and computed the corresponding 
$\langle \rho_b \rangle$ and ${+\sigma_{b}}/{-\sigma_{b}}$, analogously to the previous case \citep[e.g.,][]{EfrTib93}.

Usually, the $\pm \sigma_b$ interval dominates over $\pm \sigma_e$, but a more accurate 
estimation of the confidence interval for the correlation of the parent population is given 
by quadratically adding  $\sqrt (\sigma_{e}^2+\sigma_{b}^2)$, which, 
after reconversion gives us the best estimate for the $68.2\%$ confidence interval:
 $\langle \rho \rangle^{+\sigma_{+}}_{-\sigma_{-}}$.

\item{\bf Partial correlation}

When dealing with multivariate variables, we constantly face the same question: given three variables A, B, and C, 
is the correlation observed between the two variables A and B ($\rho_{\,AB}$) due to the association of A and B with 
the third variable, C, and not to the direct association between them? Using the partial correlation, we may eliminate the suspected 
effects of this third variable C on the relation between A and B by making C somehow constant, denoted by $\rho_{\,AB.C}$. 
This is achieved by

\begin{equation}
\rho_{\,AB.C}=\frac{{\rho}_{\,AB}-{\rho}_{\,AC}\:{\rho}_{\,BC}}
{\sqrt{(1-{{\rho}_{\,AC}}^2)(1-{{\rho}_{\,BC}}^2)}}
\label{eq:rho_partial}
.\end{equation}

If $\rho_{\,AB.C}$ does not differ significantly from $\rho_{\,AB}$ , then the association between A and B is most probably 
direct and not masked by their correlations with C. 
To take into account the effects of data error-bars, we have used the same process as previously described for the 
Spearman-rank correlation.
The significance of $\rho_{\,AB.C}$ is estimated analogously to the case of $\rho$ , but with one less degree of freedom, 
that is, using $t=\rho_{\,AB.C}\,[(n-3)/(1-\rho_{\,AB.C}^2)]^{1/2}$ --- see, for instance, \cite{Mac82}.

\end{itemize}
 

\subsection{Assessing the risk of missing existing correlations}
\label{sec:risk}

Searching for correlations means testing a {\it \textup{null hypothesis}} ($H_0$), that is, variables $X$ and $Y$ do not 
show evidence for correlation, against an {\it \textup{alternative hypothesis}} ($H_1$), where variables $X$ and $Y$  
show evidence for correlation. This evidence is measured by the significance level ($SL$), 
confidence level ($CL=1-SL$), or p-value ($=SL$). The threshold p-value below which we reject $H_0$ 
is called $\alpha$. The lower the significance, the lower the probability 
of incorrectly rejecting the {\it \textup{null hypothesis}} (type I error). To diminish the odds of finding false correlations, we 
simply adopted a  stricter criterion for rejecting $H_0$, for
instance, instead of accepting an $\alpha=1\%$ probability of 
falsely finding a correlation, we accept only $\alpha=0.1\%$, or even lower. The drawback is that this increases 
the chances of not detecting correlations that might in fact exist, meaning that we increase the {\it risk}, $\beta$, of 
missing some real correlations (type II error), or decrease the $(1-\beta)$ power
to find them. 
If we wish to diminish this risk, we must either increase 
the $\alpha$ threshold with which we reject the {\it \textup{null hypothesis}} or increase the sampling. 
We have analyzed the sample size requirements, $n$, to detect  a correlation at some 
given significance level (higher or equal to $\alpha$) and to only miss a $\beta$ proportion. 
Since we wish to compare the case where the correlation is indeed $\rho \neq 0 $ with the case of 
$\rho =0$, from \cite{Modarres1996} and \cite{BonettWright2000} we obtain

\begin{equation}
\label{eq:n_exact}
\sqrt{n(n-3)}= \frac{4(z_{1-\alpha/2}+z_{1-\beta})^2(1+\rho^2/2)}{\big(\ln{\frac{1-\rho}{1+\rho}}\big)^2}
,\end{equation}

\noindent
where $z_{1-\alpha/2}$, and $z_{1-\beta}$ are the quantiles of a normal distribution for our threshold $\alpha/2$ --- 
since we are using two-tailed tests --- and risk $\beta$, respectively, from which we compute the 
smallest integer not lower than $n$, that is, $\lceil n \rceil$, ceiling of $n$.
For example, to see evidence for a correlation  at a $2.5\,\sigma$ level, 
that is, $\alpha=0.0124$ and $z_{1-\alpha/2}=2.5$, 
in a sample extracted from a parent 
population which we knew to possess $\rho=0.6$
accepting only a $\beta=0.1$ probability (risk) of missing it, that is, $z_{1-\beta}=1.2816$, we 
would need a sample of $n\geqslant39$. If we accept a higher risk of missing the detection, 
for example, $\beta=0.2$ ($z_{1-\beta}=0.8416$),  we would need only $n\geqslant31$. 


\subsection{Degraded correlations}
\label{sec:rho_degradation}

When taking into account the data error-bars, the possible correlations between the $X$ and $Y$ 
variables will diminish or degrade. That is, taking each data-point not as an exact value but as a Gaussian probability distribution 
centered on the measured value diminishes the possible association strength between the two variables. 
The larger the error-bars, the stronger this effect. 

The error-bars of surface colors of Centaurs and KBOs tend to become larger with increasing wavelength 
(e.g., from visible $B-V$ to the near-infrared $H-K$). 
It might be possible that two different pairs of color indexes, $X_1$ vs. $Y_1$ and $X_2$ vs. $Y_2$, 
correlate with equal strength (e.g., $B-V$ vs. $V-R$ and $V-R$ vs. $R-I$) --- if we precisely knew the full population, for instance. However, not only are we estimating 
the populations' $\rho$s from a limited sample, but if the error-bars of $(X_2, Y_2)$ are larger 
than those of $(X_1, Y_1)$ we might also obtain a $|\rho_2| < |\rho_1|$ or even lose evidence for correlation on 
$(X_2, Y_2)$. To analyze this possibility, we performed the following study:

\begin{enumerate}

\item We generated a parent population of 1000 $(X,Y)$ pairs with a desired correlation value.
Each variable follows a Gaussian distribution, and their values are renormalized into the 
interval $[0,1]$ (the mean will become $0.5$) by two distinct methods: a) the lowest value 
will correspond to zero, the highest to one; b) the mean value minus 
$3\,\sigma$ will correspond to zero, the mean value plus $3\,\sigma$  to one 
(with this renormalization some points may fall outside the $[0,1]$ interval).

\item We extracted from that parent population a sample of $n$ elements. We added Gaussian error bars 
to the $Y$ elements and generated 1000 new samples from the $n$ elements 
by the same process as we used for the real data-samples (see Sect. \ref{sec:stat_tests}). 
The approximate sampling error given by \cite{BonettWright2000},

\begin{equation}
\sigma^2_{\zeta}\approx(1+\rho^2/2)/(n-3)
\label{eq:sigma_zeta}
,\end{equation}

was added quadratically to the dispersion of the Fisher-transformed simulations 
--- instead of using bootstraps extractions --- before 
reconverting them to obtain a final $\langle \rho \rangle^{+\sigma_{+}}_{-\sigma_{-}}$.

\item We iterated the two previous steps for $\rho=0.8, 0.7, 0.6,$ and $0.5$, each with 
samplings of $n=15, 20, 30, 50, 75,$ and $100$, and varyed the observational error-bars 
from $\sigma_{obs}\equiv\sigma_{o}=0.00$ to $0.30$ in steps of $0.02$. 

\item We repeated these cycles with $X$ and $Y$ following a random uniform distribution instead 
of a Gaussian.

\end{enumerate}


First, the simulations show that drawing the sample from a Gaussian distribution or from a 
uniform distribution, either renormalized into the interval $[0,1]$ by the methods 1a) or 1b), has little to 
no influence on the degradation of $\rho$ values and their error-bars, with increasing 
simulated observational errors $\sigma_{o}$ . Variations are on the order of 0.01 
or lower. Hereafter, the analysis refers to the simulations drawn from a Gaussian 
distribution that was renormalized using method 1a).

Second, we find that the effects of the (simulated) observational errors on $X$ and on $Y$, which we 
denote by $\sigma_{oX}$ and $\sigma_{oY}$, are equivalent to performing the simulations 
with additional  observational errors only on the $X$ variable of 
$\sigma_{o}=\sigma_{oX}+\sigma_{oY}$, adding no errors to the $Y$ variable. For example, having a 
series of both $X$ and $Y$ values with error-bars of $0.1$ has an equivalent effect on the correlation 
values as having a series of $X$ values with error-bars of $0.2$ and $Y$ values with no errors. 
This result is most convenient and simplifies the rest of our analysis. 
 
Third, we find, empirically, that for each initial $\rho$ value its degradation with increasing 
observational errors $\sigma_{o}$ is independent of the sample size $n$. 
The estimated $68.2\%$ error interval on the measured 
$\langle \rho \rangle^{+\sigma_{+}}_{-\sigma_{-}}$ changes with $n$ because it is largely dependent on it
(see Eq. \ref{eq:sigma_zeta}), but the diminishing of $\langle \rho \rangle$ shows no evident dependence on $n$.

Using a Levenberg-Marquardt algorithm \citep{Levenberg1944, Marquardt1963}, 
we find that third-degree polynomials fitted with great 
accuracy (all cases with residuals $rms$ $<0.0063$) the detectable $\rho$ functions denoted as $\rho^*=f(\rho,\sigma_{o})$:

\begin{equation}
\label{eq:rho_degradation}
\rho^*= 
\left\{ 
\renewcommand{\arraystretch}{1.4}
 \begin{array}{l l}
1.000-0.950\sigma_o-10.544\sigma_o^2+25.349\sigma_o^3& : \rho=1.0 \\
0.900-0.497\sigma_o-8.668\sigma_o^2+18.387\sigma_o^3 & : \rho=0.9 \\
0.800-0.309\sigma_o-7.255\sigma_o^2+14.435\sigma_o^3 & : \rho=0.8 \\ 
0.700-0.176\sigma_o-6.373\sigma_o^2+12.299\sigma_o^3 & : \rho=0.7 \\ 
0.600-0.129\sigma_o-4.322\sigma_o^2+7.461\sigma_o^3 & : \rho=0.6 \\
0.500-0.0513\sigma_o-2.817\sigma_o^2+4.373\sigma_o^3 & : \rho=0.5 \\
\end{array} 
\right . 
.\end{equation}

An immediate consequence of the $\rho$ degradation with larger $\sigma_o$s is the magnitude decrease 
of the detectable $\rho^*$. That is, even if we had a parent population such that $X=Y$, hence $\rho=1$, 
the fact that a sample has error-bars will impose a limit on the detectable 
correlation, regardless of the sample size $n$. 
In practice, from the correlation analysis point of view, 
this means that it is very important to diminish 
the observational errors, but rather pointless to enlarge the sampling with the same or larger error-bars 
if the elements in the sample are already representative of the whole population, of course, and 
the significance level of the detection has reached the threshold we desire.

Another consequence is that increasing   
$\beta$ increases the risk of missing an existing correlation. Therefore, to keep $\beta$ constant 
(for a theoretically detectable correlation), we need to increase the sampling $n$. 
We return to the example in Sect. \ref{sec:risk}. We know the parent 
population has $\rho=0.6$ and we need $n\geqslant31$ to detect evidence for a correlation at a $2.5\,\sigma$ level
with a risk $\beta=0.2$ of missing it. Suppose now that each $X$ data-point has an observational error 
$\sigma_{oX}$=0.1 and each $Y$ data-point also has $\sigma_{oY}$=0.1. As we saw earlier, this is equivalent to 
transferring all the error-bars to the X variable, such that $\sigma_o=\sigma_{oX}+\sigma_{oY}=0.2$, and to assume 
Y with no errors. 
(Note that we take $X$ and $Y$ as normalized to $[0,1]$ so this error represents $10\%$ of the 
full range of values on $X$ and $10\%$ on $Y$.) From Eq. \ref{eq:rho_degradation} we see that with such observational 
errors the detectable correlation drops to $\rho^*=0.46$, and now, from Eq. \ref{eq:n_exact}, to detect evidence 
for a correlation at the same $2.5\,\sigma$ 
significance level, we will need $n\geqslant53$ data-points, that is, $70\%$ more data-points 
($n\geqslant67$ for $\beta=0.1$). 
In this work, we are able to quantitatively predict the sample size needed for the cases studied below.


\subsection{Correction for the false-discovery rate of the correlations}
\label{sec:rho_fdr}

Since for each sample or subsample we performed about $m=150$ correlation tests, the chances of finding evidence for a correlation 
increase, even if there is no correlation at all \citep[see][for a review]{Miller1981}. For example, suppose we reject the {\it \textup{null hypothesis}} 
of no correlation using a threshold $\alpha=0.0124$ ($2.5\,\sigma$ level). There is a $1.24\%$ probability of obtaining 
evidence for a correlation just by chance, even if it is not there. This means that performing 150 tests will probably lead to one or two false discoveries.
To solve this problem, we used the false-discovery rate (FDR) procedure as proposed by \cite{BenjaminiHochberg1995}. 
That is, instead of ensuring that the probability of erroneously rejecting even one of the null hypothesis is always lower than a certain $\alpha$ 
--- using the Bonferroni procedure of rejecting $H_0$ only when its p-value is $\leqslant \alpha/m$ \citep{Bonferroni1935}---, 
we ensure that the proportion of incorrectly rejected $H_0$'s, or false discoveries, is $\alpha$. 

The Benjamini-Hochberg procedure works as follows:

\begin{enumerate}

\item Order the $m$ p-values from all tests as $p_1 \leqslant p_2 \leqslant ... \leqslant p_m$; 

\item find the largest integer $k$ such that $k=max \{ i: p_i\leqslant \frac{i}{m} \alpha \},$
 
\item reject all the null hypotheses whose p-value is $\leqslant p_k$.
 
\end{enumerate}

\noindent
This procedure ensures that FDR $\leqslant \alpha$. That is, if we use $\alpha=0.0124$ ($2.5\,\sigma$) at most $1.24\%$ of all the postive results (i.e., all the rejected null
hypotheses) may be false positives. For example, if we simultaneously
test 150 hypotheses and reject fewer than 30 of them, or in other words, 
we have fewer than 30 discoveries, it is unlikely that we make even one false discovery (at $2.5\,\sigma$, naturally). 


\section{Reddening line}
\cite{2001A&A...380..347D} introduced the concept of drawing a line in the color-color plots indicating the location of objects 
that probably possess linear reflectivity spectra. 
As discussed in detail later by \cite{2004A&A...417.1145D}, a linear reflectivity spectrum does not result in a linear relation between colors, and vice versa. 
This line has been widely used to infer the linearity or non-linearity of object spectra.
However, each telescope has its own set of filters, and the differences between the effective central wavelength for the same filters among the 
different sets has a significant effect on the location of this reddening line. This effect can be strong for the visible $BVR$ colors, but it is not very 
significant for the $VRI$ bands.

\begin{table} 
\label{tab:filters}
\renewcommand{\arraystretch}{1.3}
\renewcommand{\tabcolsep}{9pt}
\caption{\label{tab:filters} Central wavelengths in {\AA} of different telescope filters} 
\centering 
\begin{tabular}{ccccc} 
\hline\hline 
                     & \multicolumn{4}{c}{Filters} \\
Telescope  & $B$ & $V$ & $R$ & $I$ \\
\hline
8.2m ESO-VLT  & 4237 & 5481 & 6480 & 7935 \\
3.6m ESO-NTT & 4212 & 5442 & 6416 & 7950 \\
10m Keck & 4377 & 5473 & 6417 & 7599 \\
8.2m Subaru & 4400 & 5500 & 6600 & 8050 \\
3.6m CFHT & 4312 & 5374 & 6581 & 8223 \\
\hline
Average & 4307.6 & 5454 & 6498.8 & 8016.2 \\
\hline
\end{tabular}
\end{table}

\begin{figure} 
\centering 
\resizebox{7.6cm}{!}{\includegraphics{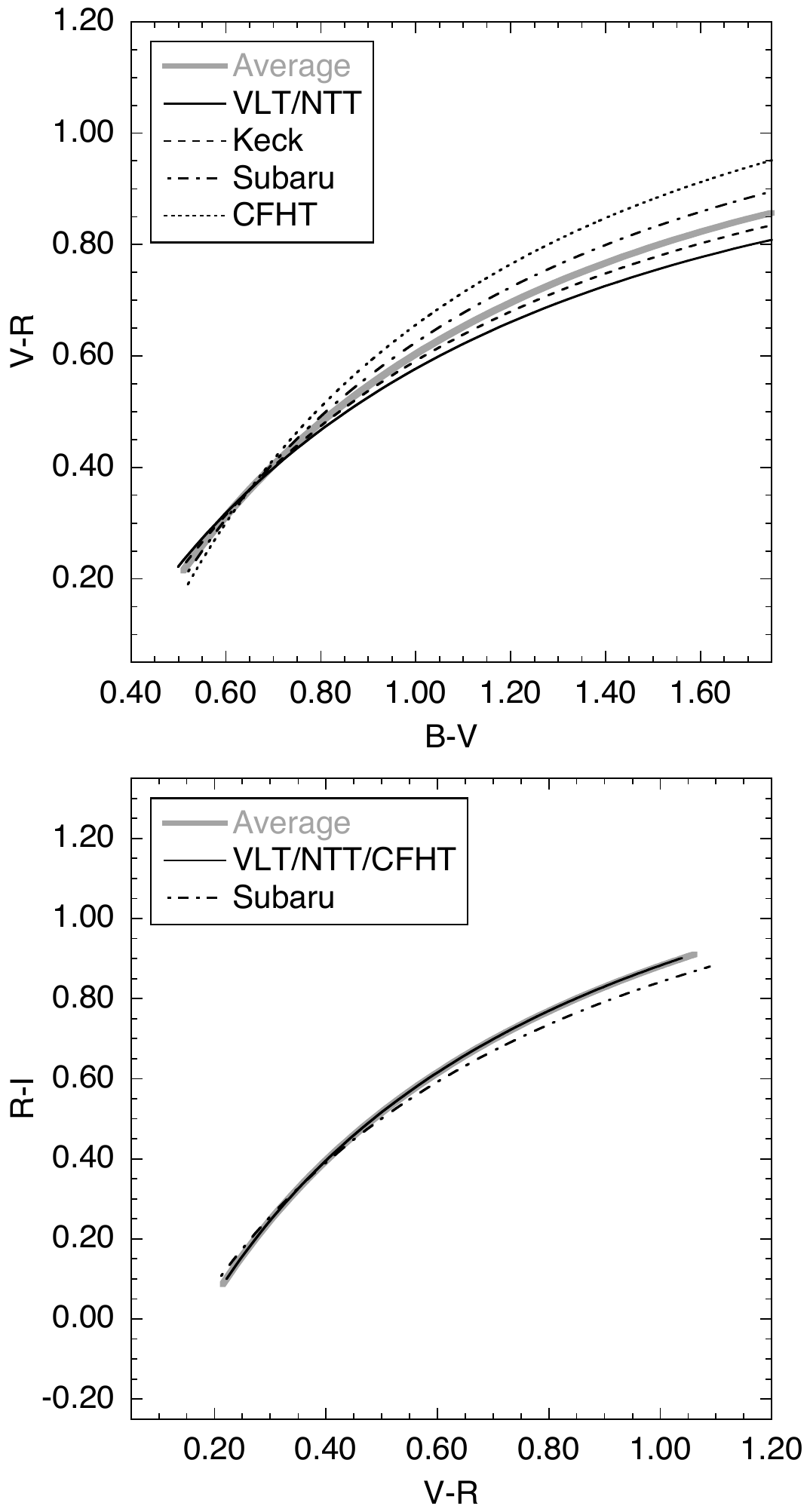}} 
\caption{$(B-V)$ vs. $(V-R)$ --- top --- and $(V-R)$ vs. $(R-I)$ --- bottom --- reddening lines obtained with the most 
frequently used sets of $BVRI$ filters in our database and the computed average reddening line. For the $(B-V)$ vs. $(V-R)$ case different filters have a strong effect on the location of the reddening line. 
For the $(V-R)$ vs. $(R-I)$ case the effect is almost negligible. 
}
\label{fig:reddening_line}
\end{figure}

We have computed the reddening line for several of the most frequently
used telescopes or instruments for KBO observations and analyzed their differences. 
As seen in Fig. \ref{fig:reddening_line}, the reddening line obtained for the Mould-$BVR$ filters of the CFHT12k camera of
the 3.6m CFHT telescope 
is higher, and the reddening lines 
obtained for the ESO FORS1 camera of the 8.2m VLT telescope or
the ESO SuSi2 camera of the 3.6m NTT telescope are the lowest. The redenings for the LRIS camera on the
10m Keck telescope and the FOCAS cameras on the 8.2m Subaru telescope are intermediate. 
Therefore, interpretations based on a precise location of this line should be taken with caution. 
We have used the solar colors $(B-V)_{\odot}=0.65$, $(V-R)_{\odot}=0.36$, and $(V-I)_{\odot}=0.70$ from \cite{Ramirez+2012}. 
Small changes in the solar colors do not affect the reddening line. 

\section{Data analysis}
\label{sec:dataanalysis}

This work is focused on $BVRI$ surface information or properties of Centaurs and KBOs. 
We have analyzed the correlations between the color indexes $B-V$, $V-R$, $R-I$, $V-I$, $B-I$, and $B-R$, 
and the spectral gradient, $Grt$ --- also known as the slope parameter or reddening $S$, which expresses the reflectivity spectrum 
variation in percent of reddening per 1000$\,\AA$ \citep{HaiDel02} ---,
and the correlations between these indexes and the absolute $R$-magnitude not corrected for phase effects, 
that is, $H_R(\alpha)\equiv R(1,1,\alpha)$. Correlations between these properties and the orbital parameters perihelion, $q$, aphelion, $Q$, orbital inclination relative to the ecliptic, $i$, orbital eccentricity, $e$, and semi-major axis, $a$, have also been studied. 
For some families we have also analyzed the correlation with the Tisserand parameter relative 
to Neptune, $T_N$, the Tisserand parameter relative to Jupiter, $T_J$, the mean random impact speed velocity, $\upsilon_{rms}$, 
the orbital excitation, $\varepsilon$, and \"Opik's $\psi$. These parameters are explained below when used. 
As mentioned in Sect. \ref{sec:data_sample}, we relied entirely on the published error bars, although for multiple measurements we checked for deviant points before carrying out the weighted averaging. 
Possible over- or underestimated errors for some data-points, most probable in the case of single measurements, 
are not expected to have any statistical influence on the methods we used and developed, unless if generalized.

Each dynamical family, or group of families, was subject to 84 tests for correlation. The exceptions were 
i) Centaurs, which were subjected to 92 tests due to the inclusion of $T_J$ in the tests; 
ii) classical KBOs, which were subjected to 100 tests, given the inclusion of $\upsilon_{rms}$, $\varepsilon$, and $\psi$; 
and iii) resonant objects were only subjected to 76 tests, since correlations with $T_N$ were not tested. 
These numbers are extremely important to correct for the false-discovery rate (FDR) as described in Sect. \ref{sec:rho_degradation}. 

We discuss the correlations for each dynamically family separately from Sects. \ref{sec:centaurs} to \ref{sec:ckbos}. 
KBOs, as a whole, are discussed in Sect. \ref{sec:kbos}, binary
or multiple objects in Sect. \ref{sec:binaries}, and all objects --- 
excluding Haumea family and retrogrades --- are discussed in Sect. \ref{sec:all}. 
For each case we also comment on the sample size and the consequent risk of missing correlations. 
Strictly speaking, this discussion is valid for a characterization of these families and groups under the assumption that 
the data are an unbiased sample of the parent population. The statistical tools, however, are general and can be applied to other 
classification systems.

\begin{figure*} 
\centering
\resizebox{16.7cm}{!}{\includegraphics{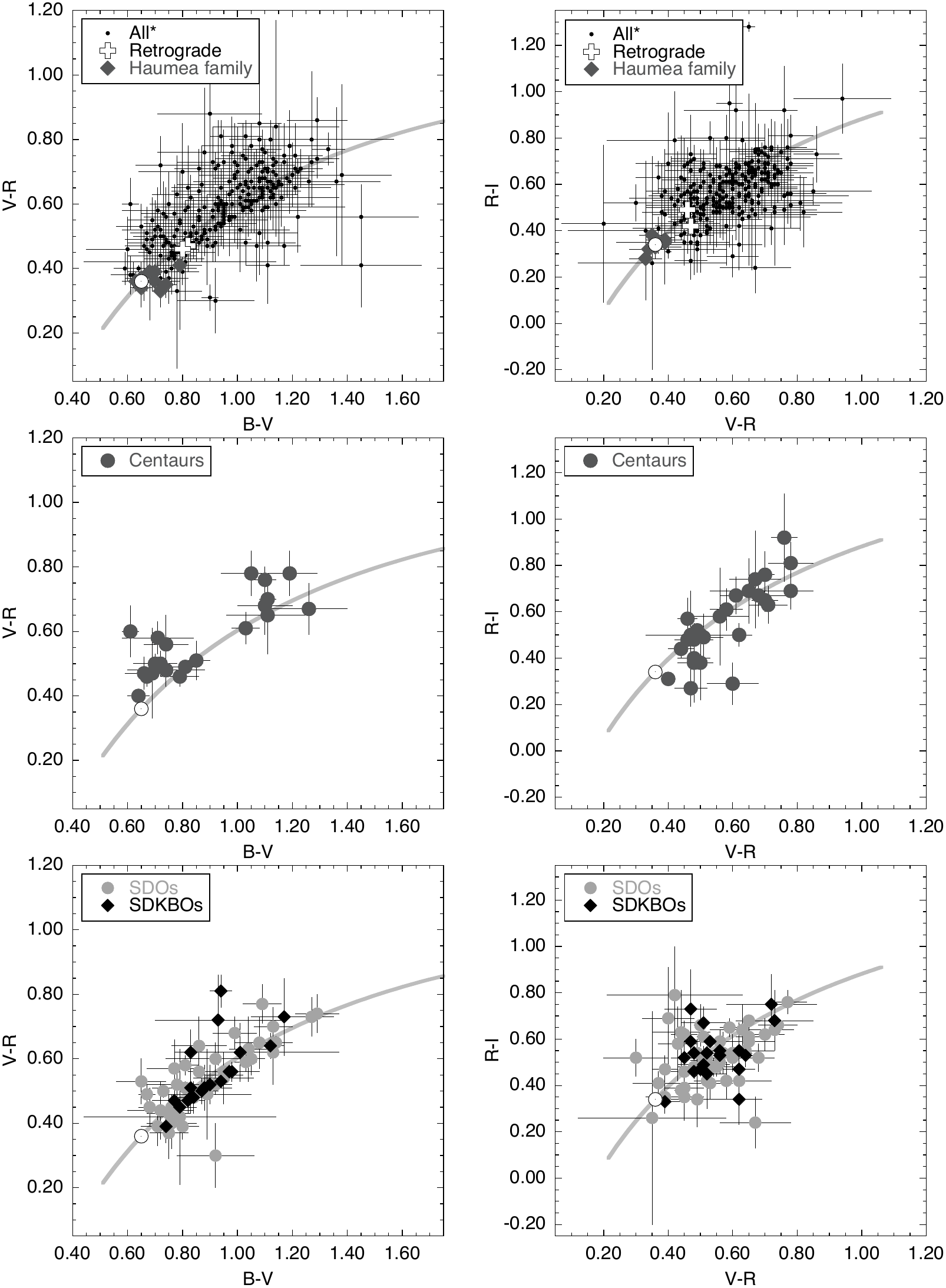}} 
\caption{$(B-V)$ vs. $(V-R)$ plots --- left column --- and $(V-R)$ vs. $(R-I)$ plots --- right column --- including the reddening line. 
Solar colors are indicated with the $\odot$ symbol. 
Top panels:  All objects, Haumea family objects, and retrograde objects. 
Middle panels: Centaurs. 
Bottom panels: Scattered disk objects (SDOs), and scattered or detached Kuiper Belt objects (SDKBOs). 
The stronger correlation between $(B-V)$ and $(V-R)$ colors 
than between $(V-R)$ and $(R-I) $ is clear. Apart from the solar-like colors of Haumea family objects and the two color groups of Centaurs 
in the $(B-V)$ vs. $(V-R)$ there is no apparent relevant difference between the color distribution of each family of objects. 
}
\label{fig:bv_vr_ri_plots1}
\end{figure*}



\onllongtab{

\begin{longtab}

\begin{longtable}{r c l r r r c l r r l l }

\caption{\label{tab:correlations} Significant Spearman $\rho$ for each family and group of families}\\ 

\hline\hline
\multicolumn{7}{c}{} & \multicolumn{3}{c}{FDR correction\tablefootmark{a}}  & & \\
\cline{8-10} \\
\multicolumn{1}{c}{$X$\tablefootmark{b}} & {\it vs.} & \multicolumn{1}{c}{$Y$\tablefootmark{c}} & $  n\tablefootmark{d} $ & $ \rho^{+\sigma}_{ -\sigma}$\tablefootmark{e} & P-value\tablefootmark{f} & $ CL(\sigma)$\tablefootmark{g}  & $ i$\tablefootmark{h} & $ \frac{i}{m}\,2.5\,\sigma$\tablefootmark{i} & $ \frac{i}{m}\,3\,\sigma$\tablefootmark{j} & $\rho_{\beta\,2.5\sigma}$\tablefootmark{k} & $\rho_{\beta\,3\sigma}$\tablefootmark{l} \\
\noalign{\smallskip}
\hline
\endfirsthead

\caption{continued.}\\
\hline\hline
\multicolumn{7}{c}{} & \multicolumn{3}{c}{FDR correction\tablefootmark{a}}  & & \\
\cline{8-10} \\
\multicolumn{1}{c}{$X$\tablefootmark{b}} & {\it vs.} & \multicolumn{1}{c}{$Y$\tablefootmark{c}} & $  n\tablefootmark{d} $ & $ \rho^{+\sigma}_{ -\sigma}$\tablefootmark{e} & P-value\tablefootmark{f} & $ CL(\sigma)$\tablefootmark{g}  & $ i$\tablefootmark{h} & $ \frac{i}{m}\,2.5\,\sigma$\tablefootmark{i} & $ \frac{i}{m}\,3\,\sigma$\tablefootmark{j} & $\rho_{\beta\,2.5\sigma}$\tablefootmark{k} & $\rho_{\beta\,3\sigma}$\tablefootmark{l} \\
\noalign{\smallskip}
\hline
\endhead

\multicolumn{12}{l}{\small\sl continued on next page} \\
\hline
\endfoot

\noalign{\smallskip}
\hline\hline
\multicolumn{12}{l}{
\tablefoottext{a}{Computational steps of the correction for the false-discovery rate (see Sect. \ref{sec:rho_fdr})}
\tablefoottext{b}{$X$ variable.}
\tablefoottext{c}{$Y$ variable.}
\tablefoottext{d}{Number of data points.}
} \\ 
\multicolumn{12}{l}{
\tablefoottext{e}{Spearman-$\rho$ correlation value $\pm$ the 68.2\% uncertainty.}
\tablefoottext{f}{P-value of the correlation.}
} \\ 
\multicolumn{12}{l}{
\tablefoottext{g}{Confidence level of the correlation in $\sigma$ values of a normalized Gaussian distribution.}
} \\ 
\multicolumn{12}{l}{
\tablefoottext{h}{Rank of the correlation after ordering by increasing P-values}
} \\ 
\multicolumn{12}{l}{
\tablefoottext{i}{FDR-corrected P-value as a function of the number of tests for $2.5\,\sigma$ confidence.}
} \\ 
\multicolumn{12}{l}{
\tablefoottext{j}{FDR-corrected P-value as a function of the number of tests for $3\,\sigma$ confidence.}
} \\ 
\multicolumn{12}{l}{
\tablefoottext{k}{Maximum detectable $\rho$ at a $2.5\,\sigma$ confidence with a $\beta$ risk of failing to detect it of 10\%.}
} \\ 
\multicolumn{12}{l}{
\tablefoottext{l}{Maximum detectable $\rho$ at a $3\,\sigma$ confidence with a $\beta$ risk of failing to detect it of 10\%.}
} \\ 
\endlastfoot

\multicolumn{12}{l}{\bf Centaurs} \\
$(B-V)$ & {\it vs.} & $(B-R)$ & $  26 $ & $ 0.83^{+ 0.08 }_{ -0.13 }$ & $ <10^{-6} $ & $ 5.42 $ & $   1 $ & $ 0.000135 $ & $ 0.000029 $ & $ 0.69 $ & $ 0.75 $ \\
$Grt$ & {\it vs.} & $(B-R)$ & $  28 $ & $ 0.82^{+ 0.08 }_{ -0.13 }$ & $ <10^{-6} $ & $ 5.42 $ & $   2 $ & $ 0.000270 $ & $ 0.000059 $ & $ 0.67 $ & $ 0.73 $ \\
$Grt$ & {\it vs.} & $(V-I)$ & $  28 $ & $ 0.81^{+ 0.10 }_{ -0.20 }$ & $ <10^{-6} $ & $ 5.22 $ & $   3 $ & $ 0.000404 $ & $ 0.000088 $ & $ 0.67 $ & $ 0.73 $ \\
$Grt$ & {\it vs.} & $(V-R)$ & $  31 $ & $ 0.81^{+ 0.09 }_{ -0.15 }$ & $ <10^{-6} $ & $ >8.0 $ & $   4 $ & $ 0.000539 $ & $ 0.000117 $ & $ 0.64 $ & $ 0.71 $ \\
$(V-R)$ & {\it vs.} & $(B-R)$ & $  28 $ & $ 0.76^{+ 0.10 }_{ -0.15 }$ & $ 0.000003 $ & $ 4.69 $ & $   5 $ & $ 0.000674 $ & $ 0.000147 $ & $ 0.67 $ & $ 0.73 $ \\
$(B-I)$ & {\it vs.} & $(B-R)$ & $  19 $ & $ 0.84^{+ 0.09 }_{ -0.17 }$ & $ 0.000007 $ & $ 4.49 $ & $   6 $ & $ 0.000809 $ & $ 0.000176 $ & $ 0.77 $ & $ 0.83 $ \\
$(V-I)$ & {\it vs.} & $(B-R)$ & $  25 $ & $ 0.76^{+ 0.11 }_{ -0.19 }$ & $ 0.000009 $ & $ 4.44 $ & $   7 $ & $ 0.000943 $ & $ 0.000205 $ & $ 0.70 $ & $ 0.76 $ \\
\hline

\end{longtable}

\end{longtab}

}



\subsection{Centaurs}
\label{sec:centaurs}

Our sample has $n=33$ Centaurs. Because one of them was a retrograde object, however, we removed it and analyzed a total of $n=32$ Centaurs. 
All of their surface colors indexes strongly correlate with each other, which indicates a strong degree of predictability 
of one color when knowing another (see middle panels of 
Fig. \ref{fig:bv_vr_ri_plots1} and Table \ref{tab:correlations}). 
No correlation between colors and absolute magnitude, and between these and any orbital parameter reaches a $2.5\,\sigma$ level after correcting for FDR. 
We stress that with $n=32$ objects, the lowest correlation value we expect not to miss at a $10\%$ risk $\beta$ and at $2.5\,\sigma$ is 
$\rho_{\beta\,2.5\sigma}=0.64$. At the same $10\%$ risk, but at $3\,\sigma$ confidence level we have $\rho_{\beta\,3\sigma}=0.70$. 
In other words, with the current sampling of Centaurs there is a risk higher than $10\%$  of missing any possible strong correlation. 
Increasing the sampling of Centaurs is still very important. To have a risk lower than $10\%$ of missing any correlation stronger than 
$\approx0.5,$ we need a sample of $n\geqslant 70$ objects (more specifically, $\rho_{\beta\,2.5\sigma}\leqslant 0.46$ and 
$\rho_{\beta\,3\sigma}\leqslant 0.50$). 

\subsection{Scattered disk objects}
\label{sec:sdos}

The dynamical boundaries for the definition of scattered disk objects, or SDOs, are ill defined. Different authors used slightly different definitions.   
Following the definition we used, we have $n=62$ SDOs. Since one of them is a retrograde object and another one a Haumea family object, we 
have analyzed a total of $n=60$ SDOs.
Most color indexes correlate with each other (see bottom panels of Fig. \ref{fig:bv_vr_ri_plots1} and Table 
\ref{tab:correlations}). The exception are $(B-R)$ vs. $(R-I)$ and $(V-R)$ vs. $(R-I)$, which indicates that it is difficult to extrapolate color values, 
although interpolating them seems possible. 
There is a correlation between $H_R$ and $q$ that we interpret as a biasing effect among SDOs, meaning that for orbits with high perihelia only the brightest objects can be  detected, which
was also discussed by \cite{2012A&A...541A..92S} --- for a review on biasing effects see \cite{2008ssbn.book...59K}.
At $\beta=0.1$, with $n=61$, we have $\rho_{\beta\,2.5\sigma}=0.48$ and $\rho_{\beta\,3\sigma}=0.53$, so some medium strength correlations might still be missed. A $\sim 50\%$ increase in the sampling of SDOs, to $n=90$, would lower these values to  
$\rho_{\beta\,2.5\sigma}=0.39$ and $\rho_{\beta\,3\sigma}=0.44$. 
Note, however, that this biasing will not be removed solely by increasing the sample. 


\subsection{Scattered or detached Kuiper Belt objects}
\label{sec:SdTNOs}

According to the classification scheme we have used, our sample has $n=7$ detached KBOs (one of which is a binary object) and 
$n=15$ scattered or detached KBOs (three of which are binaries). We opted to analyze them as one group (SDKBOs). 
The data show strong correlations between 
colors that we may consider as interpolations if we convert them into a spectrophotometric spectra. To illustrate, if we have a flat spectrum, the  
colors $(B-R)$ or $(V-I)$, for example, can ben seen as mere interpolations of the $(B-I)$ color. Analogously, when in presence of a flat spectrum, 
the $(B-I)$ color, for example, could be obtained by extrapolating from $(B-R)$. 
This means that similarly as for the SDOs, consecutive colors or color extrapolations do not show evidence of correlation (see bottom panels of Fig. \ref{fig:bv_vr_ri_plots1} and Table \ref{tab:correlations}). 
For example, $(B-I)$ vs. $(B-R)$ are strongly correlated, but the wavelength $(B-R)$ is inside the $(B-I)$ wavelength range, 
while $(B-R)$ vs. $(R-I)$ 
are not correlated.
Because we are dealing with small-sampling values the $\beta=0.1$ risk shows that it is easy to miss even very strong correlations 
among SDKBOs. For $n=20$, $\rho_{\beta\,2.5\sigma}=0.76$ and $\rho_{\beta\,3\sigma}=0.82$. Clearly, it would be good to triple the sample. 


\subsection{Plutinos}
\label{sec:Plut}

Our sample has $n=49$ Plutinos, or $(3:2)$ resonants, including Pluto and four binaries. All color indexes correlate with each other 
except for $(V-R)$ vs. $(R-I)$ (see Fig. \ref{fig:bv_vr_ri_plots2} and Table \ref{tab:correlations}). 
Nonetheless, $(B-V)$ correlated with $(V-R)$, and $(B-R)$ correlates with $(V-I)$. Therefore, either color interpolation or extrapolations 
might be possible, but for the 
latter, $B$-band information would be required. No correlations between colors and orbital parameters reach a $2.5\,\sigma$ level after correcting for FDR. 
Given the non-detection of the $(V-R)$ vs. $(R-I)$ correlation with $n=42$ objects, we highlight that at $\beta=0.1$ risk 
$\rho_{\beta\,2.5\sigma}=0.56$ and $\rho_{\beta\,3\sigma}=0.63$. As for the SDOs, some medium-strength correlation might be missed here. Increasing the sample by $\sim 50\%$ is recommended. 


\subsection{Other resonants}
\label{sec:resonants}

\begin{figure*} 
\centering
\resizebox{16.7cm}{!}{\includegraphics{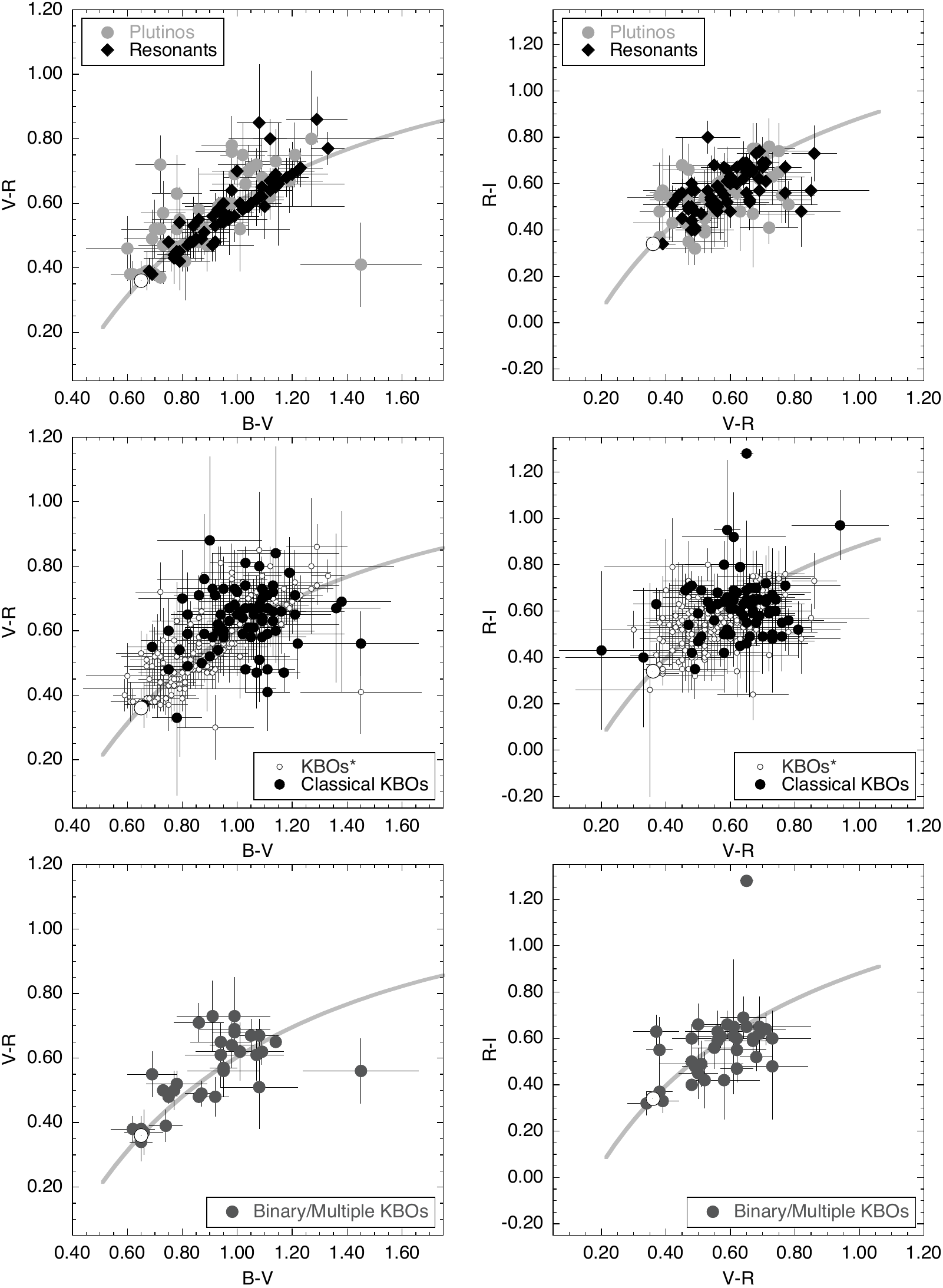}} 
\caption{$(B-V)$ vs. $(V-R)$ plots --- left column --- and $(V-R)$ vs. $(R-I)$ plots --- right column --- including the reddening line. 
Solar colors are indicated with the $\odot$ symbol. 
Top panels:  Plutinos and other resonants. Almost all non-Plutino resonants seem to lie on the reddening line in $(B-V)$ vs. $(V-R),$ but their stronger correlation value is within the error-bars of the correlations for other families.
Middle panels: Classical KBOs and all other KBOs, except Haumea family and retrogrades. 
In contrast to all other families, CKBOs do not 
exhibit a $(B-V)$ vs. $(V-R)$ correlation. The existence of many CKBOs far below the reddening 
line (in $BVR$), which indicates the presence of many concave visible spectra, suggests that they 
may exhibit absorption features in the visible that only detailed spectroscopic studies could 
verify.
Bottom panels: Binary or multiple KBOs. 
}
\label{fig:bv_vr_ri_plots2}
\end{figure*}

Our sample has $n=73$ non-Plutino resonants: 
$n=2$ $(1:1)$, or Neptune Trojans; 
$n=3$ $(5:4)$;
$n=6$ $(4:3)$;
$n=1$ $(11:8)$;
$n=1$ $(18:11)$;
$n=10$ $(5:3)$ --- 1 binary ---;
$n=1$ $(12:7)$ --- Haumea multiple system ---;
$n=1$ $(19:11)$ --- Haumea family member ---;
$n=15$ $(7:4)$ --- 1 binary ---;
$n=1$ $(9:5)$;
$n=10$ $(2:1)$ --- 3 binaries ---;
$n=3$ $(9:4)$;
$n=3$ $(7:3)$;
$n=2$ $(12:5)$;
$n=12$ $(5:2)$ --- 1 binary ---;
$n=1$ $(8:3)$ --- binary ---;
$n=2$ $(3:1)$;
$n=1$ $(11:2)$;
and $n=1$ $(11:3)$. We have removed the Haumea family members from the sample --- (136108) Haumea and (019308) 1996TO$_{66}$ --- because of
their extremely peculiar behaviors. This means that we analyzed the correlations for $n=71$ objects (see Fig. \ref{fig:bv_vr_ri_plots2} and Table 
\ref{tab:correlations}).
Absolutely all color indexes correlate with each other at a $3\,\sigma$ confidence level after the FDR correction. The correlation magnitude weakens 
to $\rho<0.6$ only for $(B-R)$ vs. $(R-I)$, $(B-V)$ vs. $(R-I)$, and $(V-R)$ vs. $(R-I)$. There is a moderate correlation between $H_R$ and $q$, suggesting that 
we have an absolute magnitude bias among these objects similarly
as for the SDOs. As for the previous cases, we detect no correlations 
between color indexes and orbital parameters. 
For $n=71$, the sample size at which the lowest $\rho$ is detected, the $\beta=0.1$ risk implies  
$\rho_{\beta\,2.5\sigma}=0.44$ and $\rho_{\beta\,3\sigma}=0.50$. Therefore, only weak or moderate correlations might be missed. 
As a whole, the other resonants do not to require a larger sampling to detect any highy relevant correlation. 
However, it is not likely that all the different resonances will possess the same surface color behaviors, and each resonance should 
be sampled such that they can be analyzed independently of the others. This still requires a large effort since even with $n=20$ objects 
one will only be able to be 90\% certain to detect very strong correlations ($\rho_{\beta\,2.5\sigma}=0.76$ and $\rho_{\beta\,3\sigma}=0.81$). 


\subsection{Classical KBOs}
\label{sec:ckbos}

Classical KBOs, also known as classicals or CKBOs, are known for their color-inclination correlation, which is not clearly detected in any 
other family os KBOs (see Fig. \ref{fig:br_i_tn_correlation} top). As for the case of other resonants, we removed the seven Haumea family objects from the classicals
sample: 
2003SQ$_{317}$, 
2003UZ$_{117}$, 
2005CB$_{79} $, 
(024835) 1995SM$_{55}$,  
(055636) 2002TX$_{300}$, 
(120178) 2003OP$_{32}$,  
and (145453) 2005RR$_{43}$. 
When plotting $B-R$ vs. orbital inclination $i,$ we have three outliers with low $B-R$ colors at very low inclination \citep[see also][]{Peixinho+2008}. 
\cite{2010AJ....140...29R} analyzed a subfamily of classical objects with semi major-axes below the nominal location of the $(3:2)$ 
resonance, calling them {\it \textup{inner classicals}}. That subfamily presented a bimodal behavior of surface colors, unlike the rest of classical objects, 
and probably a distinct origin from the primordial protoplanetary disk.  
Although we did not use the same classification scheme, the outliers 1998WV$_{24}$, and 2003YL$_{179}$ belong to that group, and  
we decided to remove them from the analysis. There is, however, no similar indication why object 2002VD$_{131}$ might be an outlier.,
but because of its extreme outlier behavior we chose to eliminate it as well. 
The final sample we analyzed has $n=117$ classical objects (see Table \ref{tab:correlations}).


\subsubsection{Correlation of color vs. color of classical KBOs}
\label{sec:ckbos_colorcolorcorr}

CKBOs evidence neither $(B-V)$ vs. $(V-R)$ nor $(V-R)$ vs. $(R-I)$ 
correlations. And with their sample size, anything stronger than $\rho \sim 0.43$ should be detected at least at a $2.5\,\sigma$ level, although within the $\rho$ error-bars we see that these correlations among KBOs$^*$, taken globally, increase when CKBOs 
are removed from the sample (see Table \ref{tab:correlations}). The middle panel of Fig. \ref{fig:bv_vr_ri_plots2} shows that
this result seems clear. 
Interestingly, there are more CKBOs in the $(B-V)$ vs. $(V-R)$ plot far below the reddening line than for any other family. 
We return to this question in Sect. \ref{sec:convex_phot_spectra}.


\subsubsection{Correlation of color vs. orbital parameters of classical KBOs}
\label{sec:coloricorr}

The trend between the colors of classical objects and their orbital inclination was first reported by \cite{2000Natur.407..979T}. 
Explicit computations of a correlation were made later by \cite{2002ApJ...566L.125T} and \cite{HaiDel02}. 

This correlation was thought to be caused by some collisional evolution mechanism, known 
as collisional resurfacing (CR) --- see \cite{1996AJ....112.2310L} --- which  seemed highly dependent on the orbital inclination because of its link 
with a stronger or weaker collisional environment.  
Different parameters have been used to analyze it, such as collisional velocity, orbital excitation, $\psi$ parameter, or simply orbital inclination.
We analyzed each one of them, also including the Tissserand parameter relative to Neptune ($T_N$).


\paragraph{{\bf Collisional velocity $\upsilon_{c}$:}}

The mean square of a KBO encounter velocity with its circular analog, that is, a non-inclined circular orbit with heliocentric distance $a$, is given by

\begin{equation}
\upsilon_{rms}=\sqrt{\frac{GM_{\odot}}{a}(e^2+i^2)}=\upsilon_K\,\sqrt{e^2+i^2}
\label{eq:vrms}
,\end{equation}

\noindent
where  $e$ represents the orbital eccentricity, $i$ stands for
the orbital inclination relative to the ecliptic, and $a$ for the semi-major axis. If $a$ is measured in AU, then $\upsilon_K\approx29.8/\sqrt{a}\;[km\,s^{-1}]$. For simplicity, we call this $\upsilon_{rms}$ simply as collisional velocity $\upsilon_{c.}$

\cite{2002ApJ...566L.125T} discussed the possible link between the color-inclination correlation and impact velocity but, given the lack of detailed modeling available at the time, they also discussed the possibility that the correlation might be a consequence of multiple subpopulations with different primordial colors and inclinations. \cite{Stern02} explored the link between KBO colors and their mean collisional velocities in more detail and also accounted for the gravitational effect of each object using $\upsilon_{coll}=\sqrt{\upsilon_{esc}^2+\upsilon_{rms}^2}$, where $\upsilon_{esc}$ is the KBO's escape velocity, calculated assuming a bulk density $\varrho=1500\;kg\,m^3$ and a geometric albedo $p=0.04$. \citeauthor{Stern02}'s Spearman-rank correlations of $(B-R)$ vs. $\upsilon_{coll}$ and $(B-V)$ vs. $\upsilon_{coll}$ were $\rho_{BR\,\upsilon_{coll}}=-0.39\;(p=0.005)$ and $\rho_{BV\,\upsilon_{coll}}=-0.44\;(p=0.0005)$, respectively, for  a total of $n=81$ KBOs. Note that these correlation values are weaker than those reported by \citeauthor{2002ApJ...566L.125T} for colors vs. inclination of classical KBOs also because \citeauthor{Stern02} analyzed all families together. 
Nonetheless, these works seemed to confirm the existence of a physically meaningful correlation between KBO colors and their mean collision velocities, making a strong case for the CR scenario. From analyzing $n=50$ CKBOs, \cite{2002AJ....124.2279D} reported $\rho_{BR\,\upsilon_{rms}}=-0.49\;(p=0.0004)$. 
Note that in spite of  \citeauthor{Stern02}'s care in also accounting for  $\upsilon_{esc}$ to the collisional velocity from our sample, we measure 
a median difference between 
$\upsilon_{coll}$ and $\upsilon_{rms}$ of only $0.02\;km\,s^{-1}$, whereas the median $\upsilon_{rms}$ is $0.55\;km\,s^{-1}$. This
is a very 
small difference given the uncertainty on the albedos and densities,
and is the reason why $\upsilon_{rms}\equiv \upsilon_{c}$ was
used more often. 
It is noteworthy that \cite{2001AJ....122.2099J}, the proponents of the CR scenario, argued against it as the primary cause of color differences, since KBOs should then also exhibit large color variations with rotation, and observations indicated otherwise. \cite{TheDor03} and \cite{The03} found additional problems with this scenario. Through a numerical analysis, they computed the kinetic energy KBOs would receive by collisions (hereafter $K_{coll}$), considering several different swarms of impactors and also different KBO distributions. Results showed correlations between $K_{coll}$ and perihelion, but also that $K_{coll}$ correlated systematically much better with eccentricity than with inclination. Observationally, there was evidence for color vs. perihelion and color vs. inclination correlations but, to date, no relevant correlations between color and eccentricity has been found. Furthermore, from these simulations, $K_{coll}$ of Plutinos (3:2 resonants) was higher than the correlation
of any other family, but observationally, their colors did not seem to differ. It is, nonetheless, true that this kind of numerical analysis has not been reanalyzed including the current understanding of the sculpting of the Kuiper Belt through migration processes. 


\paragraph{{\bf Orbital excitation $\varepsilon$:}}
\cite{HaiDel02} analyzed the color correlations with the object orbital excitation $\varepsilon$, defined as

\begin{equation}
\varepsilon=\sqrt{e^2+\sin^2\,i}
\label{eq:epsilon}
.\end{equation}

\noindent
This $\varepsilon$ parameter is related with $\upsilon_{rms}$ (see Eq. \ref{eq:vrms}), which is another estimate of the KBO velocity with respect to its circular analog. Using Pearson's correlation coefficient ($r_s$) instead of Spearman's, they obtained for $(B-R)$ vs. $\varepsilon$ of  n=13 CKBOs $r_s=-0.77$ ($p=0.002\footnote{P-value is our estimate}$), but no relevant values for any other families ($r_s<-0.1$), nor for KBOs and Centaurs taken together ($r_s=-0.21$, n=34).

\paragraph{{\bf \"Opik's $\psi$:}}
\cite{2009A&A...494..693S} analyzed the color correlations with a $\psi$ parameter, taken from the theory of \cite{Opik1976} of close encounters of a small body with a planet, given by
\begin{equation}
\psi=\frac{\frac{5}{8}\,e^2+\sin^2\,i}{a}
\label{eq:psi}
.\end{equation}

\noindent
In our context, $\psi$ relates with the average energy of the collisions experienced by a KBO. Again, the similarity between this parameter and the two previous ones is obvious. There is a subtle difference: $\sqrt{GM_{\odot}\,\psi}$ is an approximation for the KBO average encounter velocity with targets in circular orbits at the KBO instantaneous heliocentric distance on its elliptical orbit, that is, all local circular analogs. The two previous parameters consider only the impact with the global circular analog. Strangely, \cite{2009A&A...494..693S} reported for $(B-R)$ vs. $\psi$ of n=64 CKBOs $\rho=-0.02$. In our analysis we obtain $\rho_{\,(B-R)\,\psi}=-0.54 ^{+ 0.12 }_{ -0.10 }$ ($p<10^{-6}$), as roughly expected given all the previous works. 


\paragraph{{\bf Tisserand parameter $T_N$:}}
The Tisserand parameter \citep{Tisserand1896}, or invariant, has proven to be a useful tool in comet taxonomy \citep[][and references therein]{Kresak1972, Lev96}. It is a dynamical quantity kept approximately constant during a close encounter interaction between a planet and a minor 
body --- for instance, a KBO could have its orbital parameters changed by a close interaction with Neptune, and yet the Tisserand parameter value 
would not change, allowing us to keep track of the same object. 
Since KBOs are mostly under gravitational influence of Neptune, we compute it relative to this planet using

\begin{equation}
T_N=\frac{a_N}{a}+2\sqrt{\frac{a}{a_N}(1-e^2)}\,\cos{i}
\label{eq:tn}
,\end{equation}

\noindent
where $a_N=30.07\;AU$ is the semi-axis of Neptune and $a$, $e$, and $i$ are the orbital parameters of the KBO. A minor body on a circular orbit, coplanar with Neptune, with $a=a_N$ would have $T_N=3$. A KBO, being a trans-Neptunian object, will have $a>a_N$. If $T_N>3,$ then it will not cross the orbit of Neptune, whereas if $T_N<3$ means, generally, a Neptune-crosser. Usage of $T_N$ in KBO taxonomy has been suggested by \cite{Elliot+05}. Problems with the interpretation of $T_N$ values arise when high inclination values lead to a $T_N<3$, suggesting Neptune-crossing, even for impossible cases. For example, a KBO on a circular orbit with $a=37\;AU$ and $i=30^{\circ}$ would have $T_N=2.7$ and it would not cross the orbit of Neptune. Note that $T_N$ parameter also relates with the relative encounter velocity between a KBO and Neptune: $\upsilon_{c}\propto\sqrt{3-T_N}$. 

Regardless of the dynamical implications, \cite{JewPeiHsi07} reported for a total of $n=15$ CKBOs $(B-R)$ vs. $T_N$ and $(U-B)$ vs. $T_N$ correlation values of $\rho_{\,(B-R)\,T_N}=-0.75 ^{+ 0.13 }_{ -0.20 }$ ($p=0.0014$), and $\rho_{\,(U-B)\,T_N}=-0.86 ^{+ 0.08 }_{ -0.14 }$ ($p=0.00004$), respectively.


\begin{figure} 
\centering
\resizebox{7.6cm}{!}{\includegraphics{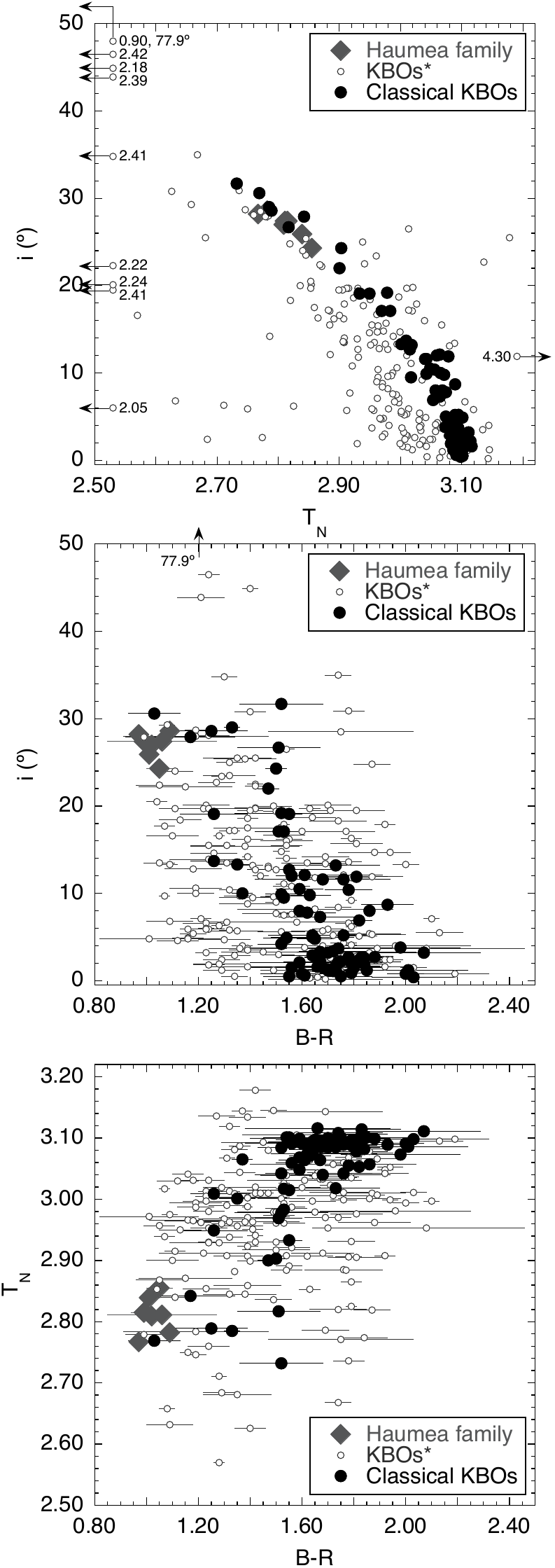}} 
\caption{Top: $(B-R)$ vs. orbital inclination plot of CKBOs and all KBOs with indication of Haumea family objects. 
The correlation for CKBOs alone is stronger than for all KBOs. The clustering of Haumea family objects had a strong 
effect on this correlation. Middle: Tisserand parameter relative to Neptune ($T_N$) vs. orbital inclination as for the above. 
$T_N$ is very strongly correlated with $i$ for  CKBOs. Bottom: $(B-R)$ vs. $T_N$ as for the above. Given the very 
strong link between $T_N$ and $i,$ statistically one cannot know if only $T_N$ or $i$ have a (potentially) physically 
meaningful relation with $(B-R)$ but none dominates the other.
}
\label{fig:br_i_tn_correlation}
\end{figure}

\paragraph{{\bf Is there only one correlation?}}

CKBOs show moderate to strong correlations, all of similar magnitudes, both with $i$, $\upsilon_{c}$, $\varepsilon$, $\psi$, and $T_N$, and a moderate to weak correlation with $q$. The confidence level for the correlation between color and eccentricity does not reach the $2.5\,\sigma$ threshold after correcting for FDR, and no relevant correlation is detected between color and semi-major axis (see Table \ref{tab:ckbo_correlations}).
From the correlations $\rho_{\,\psi \upsilon_c}= 1.00^{+0.00}_{-0.01}$, $\rho_{\,\psi \varepsilon}= 0.99^{+0.01}_{-0.01}$, 
and $\rho_{\,\varepsilon \upsilon_c }= 1.00^{+0.00}_{-0.01}$ it is clear that both $\psi$, $\varepsilon$, and $\upsilon_c$ 
will be correlated with any other variable with the same strength. Consequently, analyzing for correlations between colors 
and collisional velocity $\upsilon_c$, or orbital excitation $\varepsilon$, or  the \"Opik $\psi$ is equivalent and, 
from a statistical point of view, there is no reason to chose one over the other. Hereafter, we chose to work with 
$\upsilon_c$ since it has been used more often in other works. 


\begin{table*} 
\renewcommand{\arraystretch}{1.3}
\renewcommand{\tabcolsep}{4pt}
\caption{\label{tab:ckbo_correlations}Spearman $\rho$ correlation matrix between $(B-R)$ and orbital parameters for $n=79$ CKBOs.} 
\centering 
\begin{tabular}{cccccccccc} 
\hline\hline 
                             & $B-R$ & $T_N$ & $\psi$ & $\upsilon_{c}$ & $\varepsilon$ & $ i$ & $q$  & $e$ & $a$ \\
\hline
$B-R$                  & $\bullet$ & $ 0.50^{+0.11}_{-0.13} $ & $ -0.54^{+0.12}_{-0.10} $ & $ -0.54^{+0.12}_{-0.10} $ & $ -0.53^{+0.12}_{-0.10} $ & $ -0.50^{+0.13}_{-0.11} $ & $ 0.36^{+0.12}_{-0.13} $ & $ -0.30^{+0.13}_{-0.12} $ & $ -0.11^{+0.14}_{-0.14} $ \\
                             & & $(p=3 \cdot 10^{-6})$ & $(p<10^{-6})$ & $(p<10^{-6})$ & $(p<10^{-6})$ & $(p=2 \cdot 10^{-6})$ & $(p=0.0011)$ & $(p=0.0078)$ & $(p=0.31)$ \\
$T_N$                   & & $\bullet$ & $ -0.78^{+0.07}_{-0.05} $ & $ -0.77^{+0.07}_{-0.05} $ & $ -0.76^{+0.07}_{-0.06} $ & $ -0.85^{+0.05}_{-0.04} $ & $ 0.56^{+0.09}_{-0.10} $ & $ -0.29^{+0.11}_{-0.10} $ & $ 0.20^{+0.13}_{-0.14} $ \\
                             & & & $(p<10^{-6})$ & $(p<10^{-6})$ & $(p<10^{-6})$ & $(p<10^{-6})$ & $(p<10^{-6})$ & $(p=0.00083)$ & $(p=0.026)$ \\
$\psi$                    & & & $\bullet$ & $ 1.00^{+0.00}_{-0.01} $ & $ 0.99^{+0.01}_{-0.01} $ & $ 0.93^{+0.02}_{-0.03} $ & $ -0.66^{+0.08}_{-0.07} $ & $ 0.68^{+0.07}_{-0.08 } $ & $ 0.26^{+0.12}_{-0.13} $ \\
                             & & & & $(p<10^{-6})$ & $(p<10^{-6})$ & $(p<10^{-6})$ & $(p<10^{-6})$ & $(p<10^{-6})$ & $(p=0.034)$ \\
$\upsilon_{c}$ & & & & $\bullet$ & $ 1.00^{+0.00}_{-0.01} $ & $ 0.91^{+0.03}_{-0.03} $ & $ -0.68^{+0.09}_{-0.07} $ & $ 0.71^{+0.06}_{-0.07} $ & $ 0.28^{+0.11}_{-0.12} $ \\
                             & & & & & $(p<10^{-6})$ & $(p<10^{-6})$ & $(p<10^{-6})$ & $(p<10^{-6})$ & $(p=0.0015)$ \\
$\varepsilon$            & & & & & $\bullet$ & $ 0.91^{+0.03}_{-0.04} $ & $ -0.67^{+0.09}_{-0.07} $ & $ 0.72^{+0.06}_{-0.07} $ & $ 0.30^{+0.11}_{-0.12} $ \\
                             & & & & & & $(p<10^{-6})$ & $(p<10^{-6})$ & $(p<10^{-6})$ & $(p=0.00052)$ \\
$i$                     & & & & & & $\bullet$ & $ -0.51^{+0.09}_{-0.08} $ & $ 0.47^{+0.08}_{-0.09} $ & $ 0.13^{+0.12}_{-0.12} $ \\
                             & & & & & & & $(p<10^{-6})$ & $(p<10^{-6})$ & $(p=0.13)$ \\
$q$                        & & & & & & & $\bullet$ & $ -0.70^{+0.07}_{-0.06} $ & $ 0.04^{+0.12}_{-0.12} $ \\
                             & & & & & & & & $(p<10^{-6})$ & $(p=0.69)$ \\
$e$                        & & & & & & & & $\bullet$ & $ 0.63^{+0.07}_{-0.09} $ \\
                             & & & & & & & & & $(p<10^{-6})$ \\
$a$                       & & & & & & & & & $\bullet$ \\
                             & & & & & & & & & \\
\hline
\end{tabular}
\tablefoot{Correlation values  represented as $\rho^{+ \sigma}_{-\sigma}$. Significance p-values are two-tailed.}
\end{table*}


Now, we consider the correlations between $(B-R)$, $T_N$, $\upsilon_c$, $i$, and $q$ more closely. 
Given the mutual dependence of most of these parameters, we have carried out a partial correlation analysis to assess 
whether one of these correlation is dominant and the others are
only spurious effects with no physical relevance, but are present because of their 
dependence on the dominant one (see Sect. \ref{sec:stat_tests}).

The partial correlations between $(B-R)$, $q$, and $i$ are

\begin{eqnarray}
\left\{
\renewcommand{\arraystretch}{1.4}
\begin{array}{rrl}
\rho_{\,(B-R)i.q}=  & -0.41^{ +0.15}_{-0.13}  & (p= 0.00026; CL= 3.42 \,\sigma) \\
\rho_{\,(B-R)q.i}=  & 0.14^{ +0.14 }_{ -0.15}   & (p= 0.22; CL= 1.23\,\sigma) \\
\end{array}
\right.
,\end{eqnarray}

\noindent
showing that the correlation between $(B-R)$ colors and orbital inclination is still present and significant 
after removing the effect of the link between inclination and perihelion. Its strength, however, diminishes 
to a more moderate value. Perihelion is not a function of inclination, and the fact that $q$ and $i$ are 
correlated could be due to some dynamical evolution mechanism in the way the Kuiper Belt was sculpted, or it may simply 
be the result of a sampling bias.

The partial correlations between $(B-R)$, $T_N$, and $q$ are

\begin{eqnarray}
\left\{
\renewcommand{\arraystretch}{1.4}
\begin{array}{rrl}
\rho_{\,(B-R)T_{N}.q}=  & 0.39^{ +0.13 }_{ -0.15}   & (p= 0.00035; CL= 3.57\,\sigma) \\
\rho_{\,(B-R)q.T_{N}}=  & 0.11^{ +0.14}_{-0.14}  & (p= 0.35; CL= 0.93\,\sigma) \\
\end{array}
\right.
,\end{eqnarray}

\noindent
showing that the correlation between $(B-R)$ colors and Tisserand parameter is also still present and significant, 
although diminished to a more moderate value,  
after removing the effect of the link between $T_N$ and perihelion, analogously to the previous case.

The partial correlations between $(B-R)$, $T_N$, and $i$ are

\begin{eqnarray}
\left\{
\renewcommand{\arraystretch}{1.4}
\begin{array}{rrl}
\rho_{\,(B-R)T_{N}.i}=  & 0.17^{ +0.15 }_{ -0.15}   & (p= 0.14; CL= 1.47\,\sigma) \\
\rho_{\,(B-R)i.T_{N}}=  & -0.17^{ +0.15}_{-0.14}  & (p= 0.12; CL= 1.54\,\sigma) \\
\end{array}
\right.
,\end{eqnarray}

\noindent
which shows that none remains after eliminating the effect of the other. This should not be seen as strange given 
that by definition $T_N$ is highly dependent on $i$, presenting a $\rho_{\,T_{N}\,i=-0.85}$, so it is very hard to 
separate one from the other.

The partial correlations between $(B-R)$, $i$, and $\upsilon_c$ are

\begin{eqnarray}
\left\{
\renewcommand{\arraystretch}{1.4}
\begin{array}{rrl}
\rho_{\,(B-R)i.\upsilon_c}=  & -0.06^{ +0.15 }_{ -0.15}   & (p= 0.59; CL= 0.53\,\sigma) \\
\rho_{\,(B-R)\upsilon_c.i}=  & -0.20^{ +0.14}_{-0.13}  & (p= 0.074; CL= 1.78\,\sigma) \\
\end{array}
\right.
,\end{eqnarray}

\noindent
showing also that none remains after eliminating the effect of the other, even if the effect is stronger on 
inclination without the influence of collisional velocity than on $\upsilon_c$ without the effect of $i$. 
Analogously to the previous case, by definition the collisional velocity depends on inclination 
($\rho_{\,\upsilon_c\,i=-0.91}$), so it is hard to separate them.

Finally, the partial correlations between $(B-R)$, $T_N$, and $\upsilon_c$ are
\begin{eqnarray} 
\left\{
\renewcommand{\arraystretch}{1.4}
\begin{array}{rrl}
\rho_{\,(B-R)\upsilon_c.T_N}=  & -0.27^{ +0.14}_{-0.13}  & (p= 0.016; CL= 2.41\,\sigma) \\
\rho_{\,(B-R)T_N.\upsilon_c}=  & 0.17^{ +0.15 }_{ -0.15}   & (p= 0.14; CL= 1.48\,\sigma) \\
\end{array}
\right.
,\end{eqnarray}

\noindent
indicating that none remains without the effect of the other variable, even if the case 
of $\rho_{\,(B-R)\upsilon_c.T_N}$ is rather close to $2.5\,\sigma$.

To conclude, for CKBOs 

\begin{itemize}

\item it is statistically equivalent to use $\psi$, $\varepsilon$, or $\upsilon_c$.

\item the correlation between $(B-R)$ and perihelia may be simply an effect that arises because in the current sample of measured objects the 
perihelion is related with inclination, where inclination is
the dominating variable. The same is true if we replace the variable inclination by the 
Tisserand parameter. 

\item the correlations between $(B-R)$ and both Tisserand parameter and inclination cannot be statistically separated from the 
close link between $T_N$ and $i$. None dominates the other (see Fig. \ref{fig:br_i_tn_correlation}). 

\item although less clear than for the previous case, the correlation between $(B-R)$ and collisional velocity cannot be statistically 
separated from the dependence of $\upsilon_c$  on inclination and on $T_N$. 

\item collisional velocity $\upsilon_c$,  Tisserand parameter $T_N$, or orbital inclination $i$ may be the true physically 
relevant variables linked with $(B-R)$ surface colors. 

\item given previous studies on the lack of evidence that surface colors are affected by $\upsilon_c$, it is most likely that only 
either $T_N$ or $i$ can be the variables that truly relate with colors --- although causality cannot be assessed with this analysis ---, 
and that dynamical history plays the central role.  

\end{itemize}

It is interesting to see that when the peculiar Haumea family objects are removed from the analysis of CKBOs, 
the color-inclination correlation diminishes considerably (see Fig. \ref{fig:br_i_tn_correlation}). 
\cite{Peixinho+2008} reported $\rho=-0.70$, for $n=69$ objects, whereas now we have 
$\rho=-0.50$ for $n=79$ objects, excluding the Haumea family. 
When \cite{2005Icar..174...90D} analyzed the correlation using two different classification schemes for CKBOs, 
one using $q>35\,$AU as cutoff and another using $q>39\,$AU, they reported $\rho=-0.65$ ($n=27$) for the former and $\rho=-0.41$ 
($n=20$, and $SL<2\,\sigma$) for the latter, four of the seven objects that had $q<39\,$AU are now known to be members of the Haumea family. 
In other words, without the Haumea family, the color-inclination correlation among CKBOs is no longer a strong correlation 
but a moderate one.

It is interesting to note that CKBOs do not show evidence for
a correlation between $(B-V)$ vs. $(V-R),$ which is present in 
almost all the other more undersampled families. 
We return to this matter in Sect. \ref{sec:bvvrri_correlations}. 


\subsection{All KBOs$^*$}
\label{sec:kbos}

We have also analyzed the $n=322$ KBOs$^*$ in our sample, that is all objects except Centaurs, Haumea family objects, and retrograde orbits. 
All color indexes correlate with each other, as expected from previous works. 
However, it is interesting to note that the color-inclination correlation for all KBOs, although weaker, is within 
the error-bars of the same correlation for CKBOs. 
This suggests that the colors of most non-CKBO families of KBOs are not totally uncorrelated with inclination, otherwise the correlation would be wiped out. 
The color-inclination correlation for $n=186$ KBOs$^*$ removing the CKBOs is  
$\rho=-0.30^{+0.07}_{-0.07}$, $p=0.000042$, $CL=4.09\,\sigma$ (better than $3\,\sigma$ after FDR correction). 
We must keep in mind that if the population's $\rho$ is $-0.5$, to have only a $10\%$ risk of failing to detect it at a $2.5\,\sigma$ level 
requires a sample of $n=55$ objects, but if it is $-0.3,$ we need $n=156$. 

SDOs do have a sampling that should be enough to detect most correlations $\geqslant | \pm0.5 |,$ but show $\rho_{(B-R)i}=-0.10^{+0.15}_{-0.15}$ ($n=54$)
with a p-value $\approx 0.50$. Plutinos have a smaller sampling and show a $\rho_{(B-R)i}=-0.30^{+0.17}_{-0.15}$ ($n=45$), where the first 
correlation is below the $2.5\,\sigma$ level after correcting
for FDR. 
For non-Plutino resonants we have $\rho_{(B-R)i}=-0.26^{+0.13}_{-0.12}$ ($n=67$), but below a $2\,\sigma$ level after correcting for FDR. 
These results indicate that it is unlikely that any non-CKBO family will have a yet-to-be detected color-inclination correlation 
stronger than $-0.5$, but all may have a correlation of $\sim -0.3$ for which their sampling is too low to warrant detection. 
Clearly, the $(B-R)$ vs. $i$ trend is stronger among CKBOs, regardless of the current lower sampling of other families. 

To verify this behavior, we have simulated 1000 samples with 79 $(x, y)$ data-points correlated with $\rho=-0.50$ 
(corresponding to our CKBOs) and 189 $(x, y)$ data-points 
randomly generated, that is, with $\rho \sim 0$ (corresponding to the rest of our KBOs). The simulations show a distribution with 
$\langle \rho \rangle^{+\sigma_{+}}_{-\sigma_{-}}=-0.17^{+0.05}_{-0.05}$, which is not compatible with the observed $\rho=-0.30^{+0.07}_{-0.07}$, 
hence the weak color-inclination correlation seen among KBOs$^*$ is not simply a result of its stronger presence among CKBOs. 

Weak correlations of $\rho \sim \pm 0.2$ are detected between colors and perihelion and eccentricity. Although weak in magnitude
and only accounting for $\sim 5\%$ of the color variability --- $\rho^2$ may be roughly interpreted as the proportion of the variability 
of $Y$ that could be explained as a result of the variability of $X$ --- they are statistically significant, meaning that they are not a random effect. 


\subsection{Binary or multiple KBOs}
\label{sec:binaries}

Our sample has $n=48$ binary or multiple objects, including Pluto and Haumea, and we have also analyzed the 
primaries of these systems as a group to see if they behave differently from the non-binaries. 

Their color-color correlation does not show a different behavior from the previous cases. 
Their surface colors correlate both with $T_N$, with $\upsilon_c$, and $i$. 
A total of 26 out of these 48 objects are CKBOs, which are known to correlate with the aforementioned variables. 
Only $n=35$ binary or multiple KBOs possess measured ($B-R$) colors, which means $n=14$ CKBOs. 
Analyzing only the $n=20$ non-CKBOs, also excluding Haumea, we obtain 
$\rho=-0.49^{+0.24}_{-0.18}$, $p=0.027$, $CL=2.21\,\sigma$ (without
correcting for FDR). 
The result is not significant, and given the sample size, we have $\rho_{\beta\,2.5\sigma}=-0.76$, which means that the subsampling is too small to confirm 
the existence of $\rho_{(B-R)i} \approx -0.50$, like for CKBOs, or $\rho_{(B-R)i} \approx -0.30$ like for 
non-CKBO KBOs as discussed above. 


\subsection{All$^*$ objects}
\label{sec:all}

As indicated in Sect. \ref{sec:data_sample}, we have removed from our global analysis objects belonging to the Haumea family 
and  those with retrograde orbits. Our all$^*$ objects sample has $n=354$ objects. 
The only relevant difference between the analysis of all$^*$ objects and that of KBOs$^*$ alone is the increase 
of the correlations between absolute magnitude $H_R$ and both $q$ and $a$. These two cases may be
simply interpreted as a bias. That is, fainter objects, which are most often also smaller, are hard to observe 
at high semi-major axes or high perihelia. When we include Centaurs in the sample, which are mostly small and 
faint, this becomes even more obvious. Nonetheless, it is a known fact that there are no large Centaurs, and this 
is not a bias. 


\subsubsection{Consecutive color correlations}
\label{sec:bvvrri_correlations}

\begin{table} 
\caption{\label{tab:rho_degradation} Spearman $\rho$ for consecutive colors.} 
\centering 
\setlength{\tabcolsep}{3pt}
\renewcommand{\arraystretch}{1.4}
\begin{tabular}{ccccccc}
\hline\hline 
& &  Observed & Error & \multicolumn{2}{c}{Degraded} & Detectable \\
$X$ vs. $Y$ & $n$ &  $\rho^{+\sigma}_{-\sigma}$ & $\sigma_o$ & $\rho^*_{0.7}$ & $\rho^*_{0.6}$ & $\rho_{\beta\,2.5\sigma}$ \\
\hline
\multicolumn{7}{l}{\bf All$^*$} \\
$ (B-V)\,(V-R) $ & $ 292 $ & $ 0.54^{ +0.06 }_{ -0.07 } $ & $0.177 $ & $0.54$ &$ 0.48 $ & $ 0.22 $ \\
$ (V-R)\,(R-I) $ & $ 280 $ & $ 0.34^{ +0.07 }_{ -0.08 } $ & $0.192 $ & $0.52$ &$ 0.47 $ & $ 0.23 $ \\
\multicolumn{7}{l}{\bf KBOs$^*$} \\
$ (B-V)\,(V-R) $ & $ 266 $ & $ 0.51^{+ 0.07 }_{ -0.07 }$ & $ 0.179 $ & $0.53$ & $ 0.48 $ & $ 0.23 $ \\
$ (V-R)\,(R-I) $ & $ 252 $ & $ 0.30^{+ 0.08 }_{ -0.08 }$ & $ 0.195 $ & $0.51$ & $ 0.47 $ & $ 0.24 $ \\ 
\multicolumn{7}{l}{\bf KBOs$^*$ except CKBOs} \\
$ (B-V)\,(V-R) $ & $ 185 $ & $ 0.61^{+ 0.07 }_{ -0.08 }$ & $ 0.167 $ & $0.55$ & $ 0.49 $ & $  0.28 $ \\
$ (V-R)\,(R-I) $ & $ 175 $ & $ 0.36^{+ 0.09 }_{ -0.10 }$ & $ 0.179 $ & $0.53$ & $ 0.48 $ & $  0.29 $ \\
\multicolumn{7}{l}{\bf Classical KBOs} \\
$ (B-V)\,(V-R) $ & $ 78 $ & $ 0.15^{+ 0.15 }_{ -0.16 }$ & $ 0.234 $ & $0.47$ & $ 0.43 $ & $  0.43 $ \\
$ (V-R)\,(R-I) $ & $  76 $ & $ 0.07^{+ 0.16 }_{ -0.16 }$ & $ 0.241 $ & $0.46$ & $ 0.42 $ & $  0.43 $ \\
\multicolumn{7}{l}{\bf Other resonants} \\
$ (B-V)\,(V-R) $ & $  67 $ & $ 0.68^{+ 0.09 }_{ -0.12 }$ & $ 0.178 $ & $0.54$ & $ 0.48 $ & $  0.50 $ \\
$ (V-R)\,(R-I) $ & $  68 $ & $ 0.44^{+ 0.13 }_{ -0.15 }$ & $ 0.182 $ & $0.53$ & $ 0.48 $ & $  0.56 $ \\
\multicolumn{7}{l}{\bf SDOs} \\
$ (B-V)\,(V-R) $ & $ 53 $ & $ 0.60^{+ 0.11 }_{ -0.14 }$ & $ 0.164 $ & $0.55$ & $ 0.50 $ & $ 0.51 $ \\
$ (V-R)\,(R-I) $ & $  45 $ & $ 0.29^{+ 0.19 }_{ -0.22 }$ & $ 0.175 $ & $0.54$ & $ 0.49 $ & $ 0.56 $ \\
\multicolumn{7}{l}{\bf Plutinos} \\
$ (B-V)\,(V-R) $ & $  45 $ & $ 0.55^{+ 0.16 }_{ -0.21 }$ & $ 0.186 $ & $0.53$ & $ 0.47 $ & $  0.56 $ \\
$ (V-R)\,(R-I) $ & $  42 $ & $ 0.30^{+ 0.18 }_{ -0.21 }$ & $ 0.265 $ & $0.43$ & $ 0.40 $ & $  0.57 $ \\
\multicolumn{7}{l}{\bf Binary or multiple KBOs} \\
$ (B-V)\,(V-R) $ & $ 34 $ & $ 0.55^{+ 0.15 }_{ -0.20 }$ & $ 0.147 $ & $0.58$ & $ 0.51 $ & $ 0.63 $ \\
$ (V-R)\,(R-I) $ & $ 35 $ & $ 0.35^{+ 0.19 }_{ -0.23 }$ & $ 0.139 $ & $0.59$ & $ 0.52 $ & $ 0.65 $ \\
\multicolumn{7}{l}{\bf Centaurs} \\
$ (B-V)\,(V-R) $ & $ 26 $ & $ 0.67^{+ 0.13 }_{ -0.18 }$ & $ 0.119 $ & $0.61$ & $ 0.54 $ & $ 0.71 $ \\
$ (V-R)\,(R-I) $ & $  28 $ & $ 0.61^{+ 0.15 }_{ -0.20 }$ & $ 0.159 $ & $0.56$ & $ 0.50 $ & $ 0.69 $ \\
\multicolumn{7}{l}{\bf SDKBOs} \\
$ (B-V)\,(V-R) $ & $ 20 $ & $ 0.55^{+ 0.25 }_{ -0.40 }$ & $ 0.171 $ & $0.55$ & $ 0.49 $ & $ 0.70 $ \\
$ (V-R)\,(R-I) $ & $ 20 $ & $ 0.16^{+ 0.31 }_{ -0.35 }$ & $ 0.157 $ & $0.56$ & $ 0.50 $ & $  0.70 $ \\
\hline
\end{tabular}
\end{table}


Throughout all publications on the correlation between colors of Centaurs and KBOs, we see the tendency that with increasing wavelength of the region 
covered by the colors in question, the correlations tend to diminish. For example, in all works the correlations between $(B-V)$ vs. $(V-R)$ have been 
shown to be always 
stronger in magnitude than the correlations between $(J-H)$ vs. $(H-K)$. However, it is also a fact that the error-bars of visible colors are smaller than those 
of near-IR colors. Might this be the reason for the observed degradation of the correlation for increasing wavelength or not?

To answer this question, we have analyzed the relative median error-bar of each color normalized to the $6\,\sigma$ range of the corresponding color. For example, the 
$6\,\sigma$ range of $(B-V)$ color for all$^{*}$ is 1.036, the median error-bar is 0.08, therefore, after normalization, if $(B-V)$ ranges from 0 to 1 the 
median error-bars would be 0.077. With this procedure we can use Eq. \ref{eq:rho_degradation} and, with an assumption on the parent population $\rho$ value,  
see the effect of the error-bars on the detectable $\rho^*$ (see Sect. \ref{sec:rho_degradation})

Since the correlation of $(B-V)$ vs. $(V-R)$ usually is $\sim 0.5 - 0.6$, we assume, for simplicity, two cases: i)  
the parent population $\rho$ is 0.7 and it should be 0.7 regardless of the spectral region we examine, ii) the same as the previous, but with $\rho=0.6$. Table \ref{tab:rho_degradation} summarizes the results. 
For each $X$ vs. $Y$ pair of colors we have their own sampling $n$, observed $\rho$, and normalized median error-bars $\sigma_o$. From Eq. \ref{eq:rho_degradation} 
we compute the degraded 
$\rho_{Degraded}\equiv \rho^*$ for both 0.7 and 0.6, and from the sampling $n$ we solve Eq. \ref{eq:n_exact} to obtain the detectable $\rho$ at a $2.5\,\sigma$ level accepting only a risk $\beta=0.1$ of missing it, $\rho_{\beta\,2.5\sigma}$ (see Sect. \ref{sec:risk}). 
Statistically $\rho$ decreases with higher error-bars which are also typically associated with higher wavelengths but the variation is small. 
Observationally it decreases at a much faster rate.
Therefore, we may conclude that, globally, the $(B-V)$ vs. $(V-R)$ correlation 
is stronger than the $(V-R)$ vs. $(R-I)$ one, and it is not an effect created by larger error-bars, that is, the color-color correlations do diminish with increasing wavelength.
This result does not discard the possibility that this cannot be the case for some specific subgroups of objects, however.  

Examining these correlations for each family in more detail, we may see a similar behavior for all of them except for CKBOs. 
The sampling of Plutinos, binary or multiple KBOs, Centaurs, and SDKBOs does not warrant detection of correlations weaker than $\sim 0.6$ at a $\beta$ 
risk of $10\%$. 

Since it is clear that the $(V-R)$ vs. $(R-I)$ correlation is indeed weaker than that of $(B-V)$ vs. $(V-R)$, it is not surprising that we do not 
detect it among SDOs, Plutinos, binary or multiple KBOs, or SDKBOs. For the latter we do not even detect the $(B-V)$ vs. $(V-R)$ correlation. 
For  Centaurs, both are detected and show similar strengths. As a result of the low sampling, however, the correlation error-bars are 
too high to conclude that they are indeed similar, in contrast to what we have seen in other objects. 

The most interesting case, however, are CKBOs, since these objects show neither $(B-V)$ vs. $(V-R)$ nor $(V-R)$ vs. $(R-I)$ 
correlations (see Sect. \ref{sec:ckbos_colorcolorcorr}).
The visible spectrophotometric behavior of CKBOs, or at least one type of CKBOs, is distinct from other KBOs. We discuss it below. 


\subsubsection{Peculiar convex spectrophotometric behaviors}
\label{sec:convex_phot_spectra}

\begin{figure} 
\centering 
\resizebox{7.6cm}{!}{\includegraphics{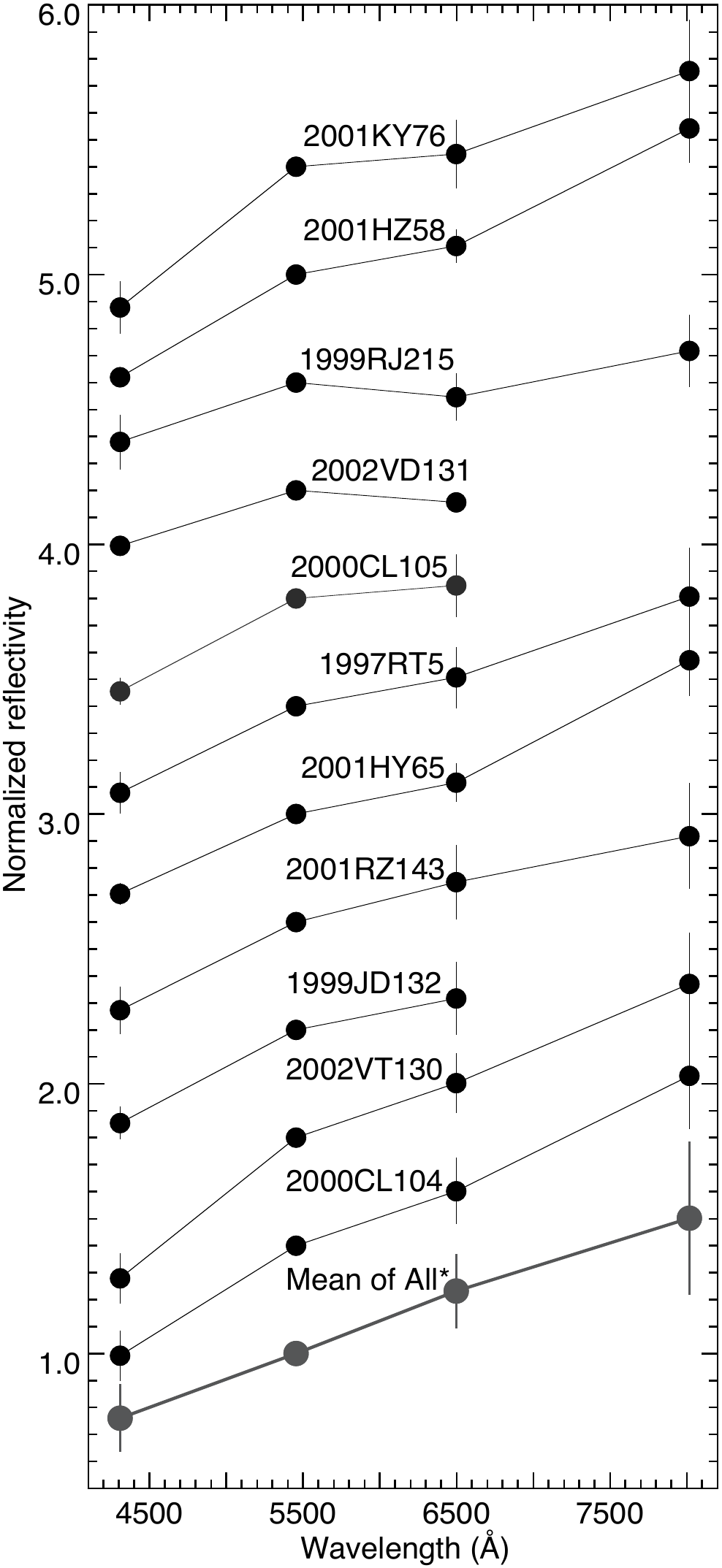}} 
\caption{Spectrophotometric reflectivity of 11 peculiar objects exhibiting a convex behavior at $\sim4500-6000\, \AA$. 
Spectra are normalized to 1 at $\lambda=5454\, \AA$ and are consecutively shifted by $0.4$ for clarity. This apparent convexity 
may be created by a concave behavior at $\sim6000-7000\, \AA,$ which suggests some absorption feature on their surfaces in this wavelength range. The spectrum resulting from the mean $B-V$, $V-R$, and $V-I$ colors,  
and their standard deviations, for all$^*$ objects is shown in gray.   
}
\label{fig:convex_phot_spectra}
\end{figure}

Surface spectra of Centaurs and KBOs tend to be straight in the visible and flatten toward the near-IR $JHK$ bands, 
most particularly when the visible slope is red, although the near-IR behavior tends to be more erratic. 
Figures \ref{fig:bv_vr_ri_plots1} and \ref{fig:bv_vr_ri_plots2} show that in $(B-V)$ vs. $(V-R)$ 
the objects tend to lie in the vicinity of the reddening line, meaning that they possess rather straight spectra in the $BVR$ wavelength range 
(pictorially: --- or $\diagup\,$), 
or above it, meaning that they possess a slightly convex spectra in that same range (pictorially: $\frown$). 
In the $(V-R)$ vs. $(R-I)$ plots we may see that objects are distributed above and below the reddening line. 
We have identified 11objects that fall far below the $BVR$ reddening line: 
2000CL$_{104}$, 2002VT$_{130}$, 2001RZ$_{143}$, 1999JD$_{132}$, 2001HY$_{65}$, 2001HZ$_{58}$, 1997RT$_{5}$, 2000CL$_{105}$,  
2002VD$_{131}$\footnote{Note that 2002VD$_{131}$ was not included in the statistical analysis of classical KBOs because of its extreme outlier behavior.} (all Classical KBOs), 2001KY$_{76}$ (Plutino), and 1999RJ$_{215}$ (SDO) --- these are all mostly classical KBOs.
Their normalized spectrophotometric reflectivity spectra are plotted in Fig. \ref{fig:convex_phot_spectra} along with the spectra resulting from 
the mean colors of all$^*$ objects (the error bars from the latter are the standard deviation of the color distribution and not the average 
error on each color). Spectra are normalized to $1$ at the center of the $V$-band filter, and each one is shifted by $0.4$ for clarity. 
The convex behavior these objects present in $BVR$ might be a mere consequence of a concave (pictorially: $\smile$) behavior in 
$VRI$, that is, the existence of an absorption band at $\sim6000-7000\, \AA$. It is tempting to associate this behavior with 
aqueous altered minerals creating absorption features in this region, as once seen with spectroscopy on $(38628)$ Huya and $(47932)$ 2000GN$_{171}$ by \cite{deBergh+04}, but the effect on visible colors is not expected to be very strong \citep[see][for a review]{2008ssbn.book..143B}.  
The object 2001RZ$_{143}$ does seem to be fully convex up to the $I$ band, and we do not possess $V-I$ colors of objects 2002VD$_{131}$, 
2000CL$_{104}$, and 1999RJ$_{215}$ to see if the spectral slope changes its convexity. Nonetheless, with the exception of 2001RZ$_{143}$, 
the other 10 aforementioned objects seem good candidates to measure for the long-sought hydration features among Centaurs and KBOs or any other 
absorption feature in the visible spectrum. 

 \subsection{Bimodal colors of small objects}

\begin{figure} 
\centering
\resizebox{7.6cm}{!}{\includegraphics{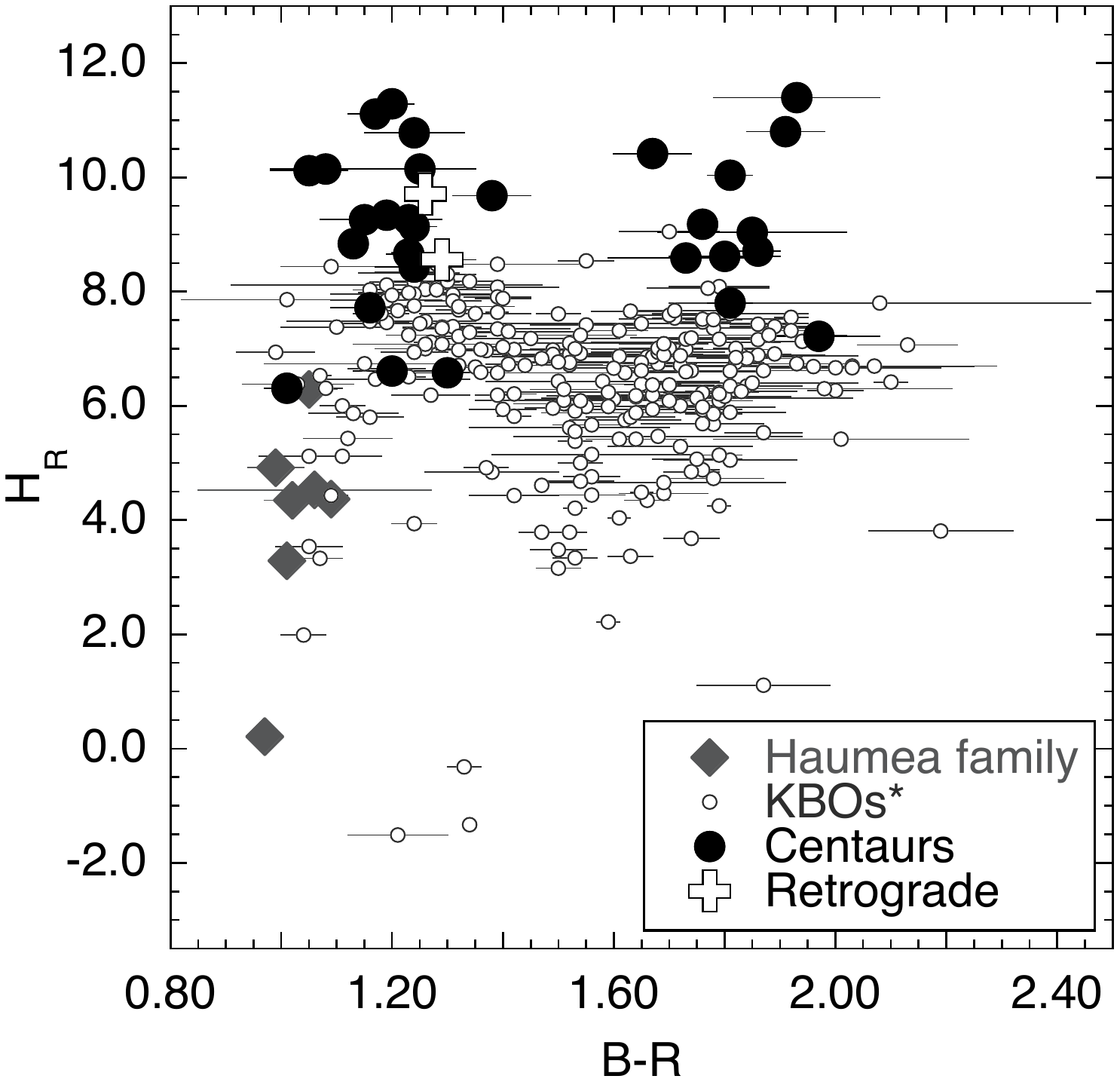}} 
\caption{$(B-R)$ colors vs. $H_R(\alpha)$ absolute magnitude (uncorrected for phase effects) 
for all objects in our database. Centaurs and KBOs are indicated with different symbols. Haumea family objects and
the two retrograde objects are also highlighted. The $\mathcal{N}$-shape reported by \cite{2012A&A...546A..86P} is clearly visible, 
and faint object bimodality is detected for $H_R(\alpha)\geqslant 6.82$, but at $2.8\sigma$ level (p-value=0.0046).
}
\label{fig:br_Hr_all}
\end{figure}

\cite{2012A&A...546A..86P} found that both Centaurs and KBOs present the so-called bimodal ($B-R$) visible color behavior at a p-value=0.001 ($3.3\,\sigma$) 
for absolute magnitudes $H_R(\alpha)\geqslant 6.8$. In a somewhat different manner, \cite{FraserBrown2012} found that 
all small objects ($H_R(\alpha)\gtrsim 6$) with perihelion $q<35$ AU are bimodal, which will therefore include Centaurs. 
With this new dataset we have carried out the same analysis as \cite{2012A&A...546A..86P}, obtaining a maximum of significance for 
the same bimodality for the $n=143$ objects with $H_R(\alpha)\geqslant 6.82$, with a p-value=0.0046 ($2.8\,\sigma$). Although the evidence for 
a bimodal behavior is very strong, we do not find a $3\,\sigma$ result with the new dataset (see Fig. \ref{fig:br_Hr_all}).

Testing the $n=112$ objects with perihelion $q<35$ AU and $H_R(\alpha)\geqslant 6.0$ from the all$^*$ sample results in a p-value=0.126 ($1.5\,\sigma$), 
which means that there is no evidence for bimodality. 

Interestingly, although the bimodality among Centaurs seems clear to the eye, when testing only for the $n=29$ Centaurs in the sample we 
obtain a p-value=0.018 ($2.4\,\sigma$).  

\section{Conclusions}

We reanalyzed the correlations of $BVRI$-band visible colors for $n=354$ Centaurs and KBOs. 
The $(B-V)$, $(V-R)$, $(R-I)$, $(B-R)$, $(V-I)$, and $(B-I)$ color indexes were carefully computed from all the 
available measurements in the literature, discarding color indexes computed from two bands taken more than two hours apart, 
and rejecting clear outliers from the averaging process. The complete sample consisted of $n=366$ objects, but we
removed from the analysis $n=2$ objects with retrograde orbits and $n=10$ Haumea family objects. Nonetheless, 
these objects were also plotted in our figures.  

\begin{itemize}

\item We used an algorithm for the Spearman-rank correlation --- also known as Spearman-$\rho$ --- taking into account not only the 
data error-bars but also the error on the estimate due to the finite sampling and corrected the significance levels using 
the {false-discovery rate} (FDR) due to the large number of tests each sample was being subjected to.

\item We analyzed for the first time the {\it risk} $\beta$ of missing any possible correlations as a result of our limited sampling and 
the sampling necessary to {\it \textup{ensure}} detection at a chosen significance level. 

\item With samples of $n<70$ objects, assumed to be unbiased extractions of the parent population, we have a $\beta>10\%$ risk of missing strong correlations, and if $n<90$ medium strength 
(moderate) correlations may still be missed. When analyzing the traditional dynamical families, only classical KBOs, as a whole, do not require larger sampling to detect presumably relevant 
correlations. If one aims to study the subfamilies of hot classical, cold classical, inner classicals, or other, then larger samples are still needed. Centaurs are particularly undersampled and should be more than doubled. Plutinos and SDOs would benefit from a $50\%$ sample increase. Other resonants, as a whole, are not undersampled, but like the Plutinos, each resonant family should be analyzed as a 
single group, and each one of them is largely undersampled. The sampling of the objects we grouped as scattered or detached KBOs would 
also need to be tripled. Note that the detection requirements
for sample-size vs. correlation strength are general and would need to be applied to 
any possible subfamily or distinct classification one decides to study. However, the samples are assumed to be unbiased. Including possible
sample biasing effects would require a more complex analysis. 

\item We studied the {\it \textup{degradation}} of a detectable correlation as a function of the data error-bars and concluded that after
a certain correlation has been detected, it is pointless to increase the sampling while maintaining the same error-bars. Only by reducing the 
observational error-bars can a possibly stronger correlation be measured.  

\item We showed that the location of so-called {\it \textup{reddening line}} in the color-color plots is quite sensitive to the central 
wavelengths of similar filters used on different telescopes. The effect is strong in the $(B-V)$ vs. $(V-R)$ plots, but weak in 
the $(V-R)$ vs. $(R-I)$ plots.

\item The correlation between $(B-V)$ and $(V-R)$ colors is stronger than the correlation between $(V-R)$ and $(R-I)$, regardless 
of the observational error-bars.

\item We analyzed the different parameters that have been used in the context of the color-inclination 
correlation of classical KBOs and the collisional resurfacing scenario using partial correlation tests. 
Statistically, it is equivalent to using the orbital excitation $\varepsilon$,  \"Opik's $\psi$, or the collisional velocity $\upsilon_c$. 
The detected color-perihelion correlation is most likely physically irrelevant and an effect of the strong relation between 
perihelion $q$ and inclination $i$ among CKBOs.
Both orbital inclination $i$ and Tisserand parameter relative to Neptune $T_N$ are the most likely physically relevant parameters 
related with the color variation of CKBOs, therefore the relation is probably related to dynamical history alone. Statistically, 
the collisional velocity cannot be separated from its dependence on inclination $i$ or $T_N$. 
Without the Haumea family, the color-inclination correlation of CKBOs is no longer strong, but moderate ($\rho\sim0.5$). 

\item Although weaker than for CKBOs, a color-inclination correlation must still exist among the non-classical KBOs. 

\item Classical KBOs are the only family that show neither $(B-V)$ vs. $(V-R)$ nor $(V-R)$ vs. $(R-I)$ correlation. 

\item We identified ten objects, only two of which are not CKBOs, that exhibit a convex spectral behavior from their $(B-V)$ and $(V-R)$ colors 
and a concave behavior from their $(V-R)$ and $(R-I)$ colors, suggesting an absorption band or feature at $\sim 6000 - 7000\, \AA$, which renders them 
good candidates for detailed spectroscopic studies.

\item We detected evidence for the $(B-R)$ color bimodality for objects with absolute magnitude $H_R(\alpha)\geqslant 6.82$ at a 
$2.8\,\sigma$ level (p-value=0.0046) --- see \cite{2012A&A...546A..86P}. 
Paradoxically, the evidence for bimodality among Centaurs reaches only $2.4\,\sigma$ --- see \cite{Peix+03L} ---, and  
we find no evidence for bimodality of objects with perihelion $q<35$ AU and $H_R(\alpha)\geqslant 6.0$ --- see \cite{FraserBrown2012}.

\end{itemize}

 We must stress that we analyzed the correlations and sampling requirements of the 
traditional dynamical families of KBOs and Centaurs as defined in Sect. \ref{sec:data_sample},
 with the aim to characterize each of them, and with the underlying statistical assumption that each sample is or will be unbiased. 
 The physical and/or chemical relevance of the dynamical classification system used was not discussed. 
 Using distinct classification systems, subsystems, or groupings should lead to distinct correlations
 as well as to distinct sampling considerations, and such works are to be encouraged. 
 
 A similar study, extended to the near-IR colors, is under preparation.


\begin{acknowledgements}
We thank the anonymous referee for the comments that helped to improve this manuscript. 
We thank Audrey Thirouin for her LaTeX ``magic''. 
NP acknowledges funding by the Gemini-Conicyt Fund, allocated to the project 
N\textsuperscript{\underline{o}} 32120036. 
\end{acknowledgements}




\onllongtab{
\begin{landscape}
\begin{scriptsize}
\setlength{\tabcolsep}{2pt}
\begin{longtable}{r l r r r r r r r r r r r r r r r r r r r p{35pt}}
\caption{\label{onlinetab:colors}Compilation of absolute magnitude and visible colors and corresponding orbital parameters }\label{tab:colors}\\
\hline\hline
 & Object & Clas\tablefootmark{a} & \multicolumn{1}{c}{$H_R$\tablefootmark{b}} & \multicolumn{1}{c}{$Grt$\tablefootmark{c}} & \multicolumn{1}{c}{$B-V$} & \multicolumn{1}{c}{$V-R$} & \multicolumn{1}{c}{$R-I$} & \multicolumn{1}{c}{$V-I$} & \multicolumn{1}{c}{$B-I$} & \multicolumn{1}{c}{$B-R$} & \multicolumn{1}{c}{$q$} & \multicolumn{1}{c}{$Q$} & \multicolumn{1}{c}{$i$} & \multicolumn{1}{c}{$e$} & \multicolumn{1}{c}{$a$} & \multicolumn{1}{c}{$\upsilon_c$\tablefootmark{d}} & \multicolumn{1}{c}{$\psi$\tablefootmark{e}} & \multicolumn{1}{c}{$\varepsilon$\tablefootmark{f}} & \multicolumn{1}{c}{$T_N$\tablefootmark{g}} & \multicolumn{1}{c}{$T_J$\tablefootmark{h}} & References \\
\hline
\endfirsthead
Table \ref{onlinetab:colors} cont'd\\
\hline\hline
 & Object & Clas\tablefootmark{a} & \multicolumn{1}{c}{$H_R$\tablefootmark{b}} & \multicolumn{1}{c}{$Grt$\tablefootmark{c}} & \multicolumn{1}{c}{$B-V$} & \multicolumn{1}{c}{$V-R$} & \multicolumn{1}{c}{$R-I$} & \multicolumn{1}{c}{$V-I$} & \multicolumn{1}{c}{$B-I$} & \multicolumn{1}{c}{$B-R$} & \multicolumn{1}{c}{$q$} & \multicolumn{1}{c}{$Q$} & \multicolumn{1}{c}{$i$} & \multicolumn{1}{c}{$e$} & \multicolumn{1}{c}{$a$} & \multicolumn{1}{c}{$\upsilon_c$\tablefootmark{d}} & \multicolumn{1}{c}{$\psi$\tablefootmark{e}} & \multicolumn{1}{c}{$\varepsilon$\tablefootmark{f}} & \multicolumn{1}{c}{$T_N$\tablefootmark{g}} & \multicolumn{1}{c}{$T_J$\tablefootmark{h}} & References \\
\hline
\endhead
\hline
\endfoot
        & 1993RO          &     Plut & $  8.48\pm 0.11 $ & $ 19.02\pm  8.66 $ & $ 0.86\pm 0.11 $ & $  0.58\pm 0.13 $ & $  0.52\pm 0.19 $ & $  1.02\pm 0.22 $ & $  1.70\pm 0.29 $ & $  1.39\pm 0.15 $ & $    31.521$ & $   47.275$ & $    3.7$ & $   0.1999$ & $   39.398$ & $   1.00$ & $   0.0007406$ & $   0.210$ & $    3.002$ & $   5.513$ &  1, 2, 3 \\ 
        & 1994ES$_{2  }$  &    cClas & $  7.53\pm 0.12 $ & $ 74.75\pm  9.01 $ & $ \text{---}          $ & $  0.94\pm 0.15 $ & $  0.97\pm 0.15 $ & $  1.91\pm 0.21 $ & $  \text{---}          $ & $  \text{---}          $ & $    40.686$ & $   50.977$ & $    1.1$ & $   0.1123$ & $   45.832$ & $   0.50$ & $   0.0001795$ & $   0.114$ & $    3.109$ & $   6.011$ &  1 \\ 
        & 1994EV$_{3  }$  &    cClas & $  7.12\pm 0.07 $ & $ 29.48\pm  7.07 $ & $ 1.09\pm 0.16 $ & $  0.58\pm 0.12 $ & $  0.80\pm 0.10 $ & $  1.47\pm 0.18 $ & $  2.54\pm 0.21 $ & $  1.73\pm 0.17 $ & $    40.903$ & $   44.690$ & $    1.7$ & $   0.0442$ & $   42.797$ & $   0.24$ & $   0.0000481$ & $   0.053$ & $    3.085$ & $   5.849$ &  1, 4, 5, 6 \\ 
        & 1994TA          &     Cent & $ 11.40\pm 0.13 $ & $ 32.61\pm  6.24 $ & $ 1.26\pm 0.14 $ & $  0.67\pm 0.08 $ & $  0.74\pm 0.21 $ & $  1.36\pm 0.29 $ & $  2.72\pm 0.41 $ & $  1.93\pm 0.15 $ & $    11.791$ & $   21.830$ & $    5.4$ & $   0.2986$ & $   16.811$ & $   2.27$ & $   0.0038409$ & $   0.313$ & $    3.210$ & $   3.725$ &  3, 7 \\ 
        & 1995DB$_{2  }$  &    cClas & $  8.11\pm 0.09 $ & $  \text{---}           $ & $ \text{---}          $ & $  \text{---}          $ & $  \text{---}          $ & $  1.63\pm 0.22 $ & $  \text{---}          $ & $  \text{---}          $ & $    40.151$ & $   53.041$ & $    4.1$ & $   0.1383$ & $   46.596$ & $   0.68$ & $   0.0003663$ & $   0.156$ & $    3.105$ & $   6.024$ &  8 \\
\hline
\end{longtable}
\tablefoot{
   \tablefoottext{a}{Dynamical class or family --- see Sect. \ref{sec:data_sample}. Binary or multiple objects are indicated with (b); retrograde orbits with (r); and Haumea collisional family objects with (h).}
   \tablefoottext{b}{R-filter absolute magnitude not corrected for phase effects. It may also be denoted by $H_R(\alpha)$ or $R(1,1,\alpha)$.}
   \tablefoottext{c}{Spectral gradient, slope parameter, or reddening $S$. It expresses the reflectivity spectrum 
variation in percent of reddening per 1000$\,\AA$ --- see \cite{HaiDel02}}
   \tablefoottext{d}{Collisional velocity. The mean square of an object's encounter velocity with its circular analog --- see Sect. \ref{sec:coloricorr}.}
   \tablefoottext{e}{\"Opik's $\psi$. Another estimate of an object's encounter velocity with its circular analog --- see Sect. \ref{sec:coloricorr}.}
   \tablefoottext{f}{Orbital excitation. It relates with collisional velocity --- see Sect. \ref{sec:coloricorr}.}
   \tablefoottext{g}{Tisserand parameter, or invariant, relative to Neptune --- see Sect. \ref{sec:coloricorr}.}
   \tablefoottext{h}{Tisserand parameter, or invariant, relative to Jupiter.}
   }
\tablebib{(1) ~\citet{1996AJ....112.2310L};
(2) \citet{2001A&A...378..653B};
(3) \citet{2000Natur.407..979T};
(4) \citet{2001Icar..152..246G};
(5) \citet{2002A&A...395..297B};
(6) \citet{2002ApJ...566L.125T};
(7) \citet{2001AJ....122.2099J};
(8) \citet{2011Icar..213..693B};
(9) \citet{1997MNRAS.290..186G};
(10) \citet{2000AJ....120..496B};
(11) \citet{2002AJ....124.2279D};
(12) \citet{1998AJ....115.1667J};
(13) \citet{2003Icar..161..181T};
(14) \citet{2001A&A...380..347D};
(15) \citet{2009A&A...494..693S};
(16) \citet{2010A&A...511A..72S};
(17) \citet{2001Icar..154..277D};
(18) \citet{2012AJ....144..169S};
(19) \citet{2004Icar..170..153P};
(20) \citet{2003ApJ...599L..49T};
(21) \citet{2006Icar..183..168G};
(22) \citet{2005Icar..174...90D};
(23) \citet{2007AJ....134.2186D};
(24) \citet{2010AJ....139.1394S};
(25) Tegler et al. http://www.physics.nau.edu/~tegler/research/survey.htm;
(26) \citet{2004A&A...421..353F};
(27) \citet{2010A&A...510A..53P};
(28) \citet{2008AJ....136.1502R};
(29) \citet{2010AJ....140...29R};
(30) \citet{1997P&SS...45.1607L};
(31) \citet{2003Icar..166..195B};
(32) \citet{2003A&A...400..369R};
(33) \citet{1997AJ....113.1893R};
(34) \citet{2002Icar..160...59R};
(35) \citet{2007AJ....133...26R};
(36) \citet{2001A&A...371..753P};
(37) \citet{1997Icar..126..212T};
(38) \citet{1999Icar..142..476B};
(39) \citet{2000A&A...356.1076H};
(40) \citet{2004A&A...415L..21B};
(41) \citet{2002A&A...392..335B};
(42) \citet{2001ApJ...548L.243F};
(43) \citet{2010A&A...511A..35A};
(44) \citet{2003Icar..162..408D};
(45) \citet{2010A&A...521A..35D};
(46) \citet{2005P&SS...53.1501D};
(47) \citet{2005A&A...437.1115D};
(48) \citet{2006ApJ...639.1238R};
(49) \citet{2009Icar..200..292B};
(50) \citet{2012A&A...546A..86P}
}
\end{scriptsize}
\end{landscape}
}


\end{document}